\documentclass[twocolumn,prl,superscriptaddress,preprintnumbers,nofootinbib]{revtex4}[11pt]

\usepackage{amsmath,amssymb,graphicx}
\graphicspath{{figs/}}
\usepackage{epsf,verbatim}
\usepackage{hyperref}
\usepackage{comment}
\usepackage{color}
\usepackage{slashed}
\usepackage{subfigure}
\usepackage[usenames,dvipsnames]{xcolor}
\usepackage{comment}
\usepackage{mdwlist, paralist}
\usepackage{rotating}
\usepackage{booktabs}
\usepackage{bm}
\usepackage{multirow}
\usepackage{topcapt}
\usepackage{scalerel}
\usepackage{cancel}

\newcommand{\beq}{\begin{equation}}
\newcommand{\eeq}{\end{equation}}
\newcommand{\bea}{\begin{eqnarray}}
\newcommand{\eea}{\end{eqnarray}}
\newcommand{\bit}{\begin{itemize}}
\newcommand{\eit}{\end{itemize}}
\newcommand{\ben}{\begin{enumerate}}
\newcommand{\een}{\end{enumerate}}

\newcommand{\alg}[1]{\begin{align} \begin{split} #1 \end{split}  \end{align}}

\renewcommand{\eqref}[1]{Eq.~(\ref{eq:#1})}

\newcommand{\figref}[1]{Figure~\ref{fig:#1}}

\newcommand{\tabref}[1]{Tab.~\ref{tab:#1}}

\newcommand{\f}{\frac}

\newcommand{\vev}[1]{ \left\langle {#1} \right\rangle }

\newcommand{\cmg}{\,\text{cm}^2/\text{g}}

\newcommand{\mkpc}{M_\odot/\text{kpc}^{3}}

\newcommand{\kms}{\text{km}/\text{s}}

\newcounter{sec}

\begin{document}
\title{Constraining Dissipative Dark Matter Self-Interactions} 

\author{Rouven Essig}
\affiliation{C.N.~Yang Institute for Theoretical Physics, Stony Brook University, Stony Brook, New York 11794}

\author{Samuel D. McDermott}
\affiliation{Fermi National Accelerator Laboratory, Center for Particle Astrophysics, Batavia, Illinois 92376}

\author{Hai-Bo Yu}
\affiliation{Department of Physics and Astronomy, University of California, Riverside, California 92521}

\author{Yi-Ming Zhong}
\thanks{ymzhong@bu.edu}
\affiliation{Physics Department, Boston University, Boston, Massachusetts 02215}

\begin{abstract} 

We study the gravothermal evolution of dark matter halos in the presence of dissipative dark matter self-interactions. Dissipative interactions are present in many particle-physics realizations of the dark-sector paradigm and can significantly accelerate the gravothermal collapse of halos compared to purely elastic dark matter self-interactions. This is the case even when the dissipative interaction timescale is longer than the free-fall time of the halo. Using a semianalytical fluid model calibrated with isolated and cosmological $N$-body simulations, we calculate the evolution of the halo properties -- including its density profile and velocity dispersion profile -- as well as the core-collapse time as a function of the particle model parameters that describe the interactions. A key property is that the inner density profile at late times becomes cuspy again. Using 18 dwarf galaxies that exhibit a corelike dark matter density profile, we derive constraints on the strength of the dissipative interactions and the energy loss per collision.

\end{abstract}

\preprint{FERMILAB-PUB-18-437-A, YITP-SB-18-21}
\maketitle

\stepcounter{sec}
{\emph {Introduction.---}\;}The elusive nature of dark matter (DM) in terrestrial experiments combined with hints for nontrivial dynamics from astrophysical systems has led to the dark sector paradigm: the DM may be connected to a plethora of hidden particles with their own interactions; see Refs.~\cite{Battaglieri:2017aum,Alexander:2016aln,Essig:2013lka} for overviews. These dark-sector interactions may modify the formation and evolution of DM halos and alter their inner structure.  Astrophysical observations can in turn provide important tests on the microscopic physics in the dark sector.

In this Letter, we explore observational consequences of a generic dark-sector model, where DM particles have both elastic and dissipative self-interactions. Self-interacting DM (SIDM) has been proposed to solve long-standing issues of the prevailing cold DM model on galactic scales; see Ref.~\cite{Tulin:2017ara} for a review. Most SIDM studies focus on the elastic scattering limit. However, in many particle physics realizations of SIDM~\cite{Kaplan:2009de,CyrRacine:2012fz,Cline:2013pca,Boddy:2014qxa,Finkbeiner:2014sja,Foot:2014uba,Boddy:2016bbu,Schutz:2014nka,Zhang:2016dck,Blennow:2016gde,Das:2017fyl}, DM particles also have dissipative collisions. We show that observations of constant DM density cores in many dwarf galaxies can be used to test dissipative DM self-interactions.  

A finite self-gravitating system has negative heat capacity, and the evolution of an SIDM halo culminates in the ``gravothermal catastrophe"~\cite{LyndenBell:1968yw}: over sufficiently long timescales, the inner core ultimately experiences gravitational collapse and a cuspy density profile reappears~\cite{Kochanek_2000}. If this were to occur, SIDM would, in fact, fail to explain the low-density cores exhibited in many dwarf and low surface brightness (LSB) galaxies~\cite{Flores:1994gz,Moore:1994yx,Gentile:2004tb,Persic:1995ru,KuziodeNaray:2007qi,deBlok:2008wp,Oh:2010ea,Oh:2015xoa}. Interestingly, if the self-interactions are exclusively elastic, halo core collapse only occurs within the age of the Universe for self-scattering cross section per unit mass $\sigma/m\gtrsim10\textup{--}50~{\rm cm^2/g}$~\cite{Vogelsberger:2012ku,Elbert:2014bma}, whereas $\sigma/m\sim \mathcal O(1)~{\rm cm^2/g}$ is sufficient to explain stellar kinematics in dwarfs~\cite{Spergel:1999mh,Firmani:2000qe,Dave:2000ar,Vogelsberger:2012ku,Rocha:2012jg,Zavala:2012us,Kamada:2016euw,Elbert:2014bma,Kaplinghat:2015aga,Valli:2017ktb,Ren:2018jpt}. However, in the presence of dissipative interactions, the gravothermal evolution of an SIDM halo can be accelerated significantly, as we will show. 

We focus on the ``mild cooling regime,'' in which the cooling timescale is longer than the free-fall time of the halo. In this case, the halo mostly stays in hydrostatic equilibrium and contracts as a whole without fragmentation, as opposed to situations with strong cooling~\cite{Fan:2013tia, Buckley:2014hja, DAmico:2017lqj,Buckley:2017ttd,Outmazgine:2018orx}. After introducing a physical model to capture the bulk cooling, we perform numerical simulations to trace the evolution of the halo and calibrate the results against both isolated and cosmological $N$-body simulations. Finally, we derive strong limits on the strength of dissipative interactions in the dark sector. In Supplemental Material, we provide additional details and results to further support our main text.

\stepcounter{sec}
{\emph {Methodology.}---\;}To understand halo evolution in the presence of dissipative interactions, we employ a semianalytical fluid model, which has been used to study globular clusters~\cite{lynden1980consequences,spitzer2014dynamical,binney2011galactic} and halos consisting of SIDM without dissipation~\cite{Gnedin:2001gd,Balberg:2001qg,Balberg:2002ue,Koda:2011yb,Pollack:2014rja}. Since this method is computationally inexpensive, we are able to scan a wide range of parameter space. Moreover, it can resolve the very inner regions of the simulated halo.

\begin{figure}[t]
  \includegraphics[width=0.48\textwidth]{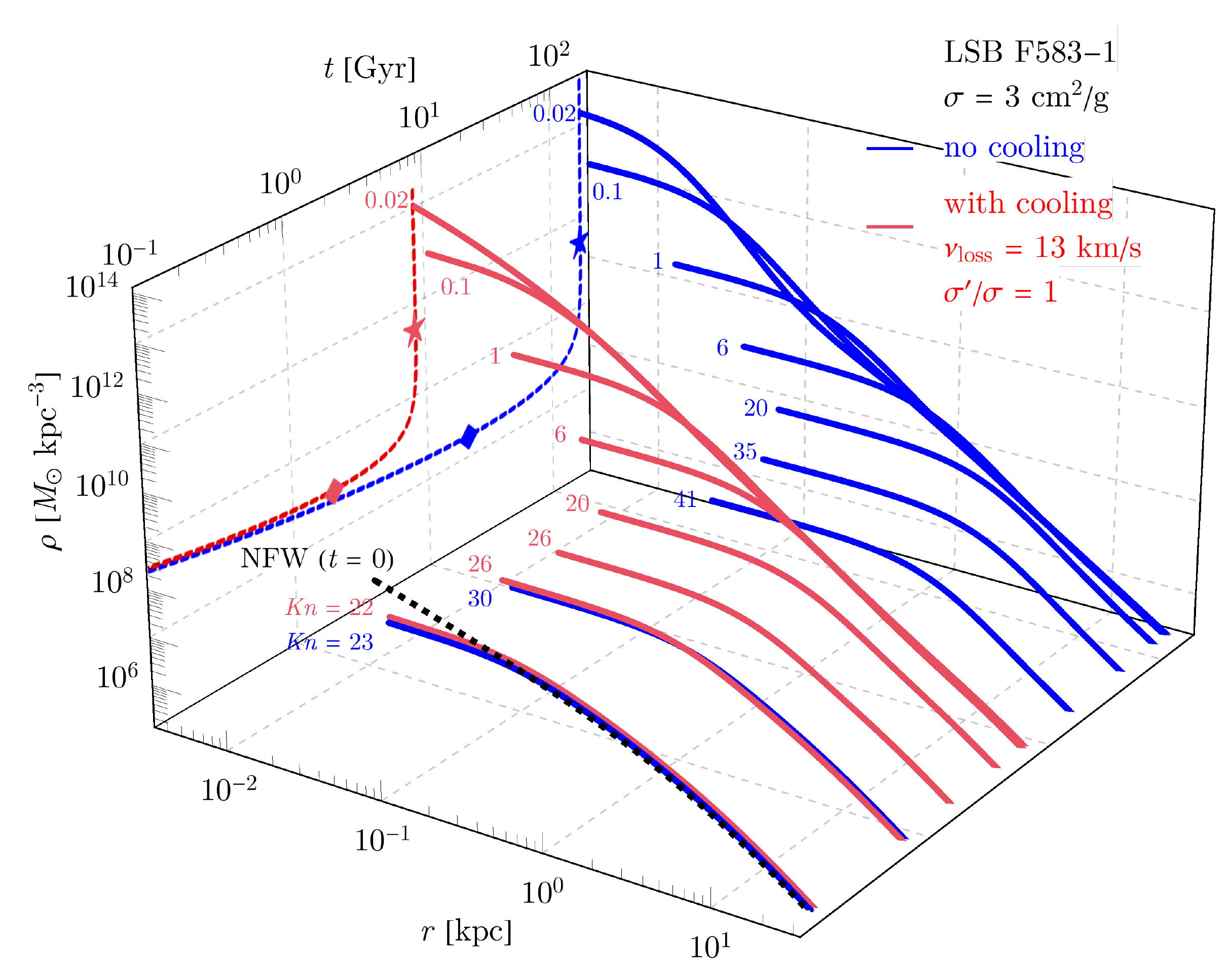}
   \caption{The evolution of the density profiles for an SIDM halo, LSB F583-1, assuming 
   purely elastic DM self-interactions (``no cooling'', blue) 
   and self-interactions with an additional dissipative interaction (``with cooling'', red). Numbers show the Knudsen number for the innermost shell at a given time. We project the evolution of the density of the innermost shell on to the $\rho\textup{--}t$ plane and mark the separation of stages $1\to2$ and $2\to3$ with diamond and star, respectively. We take $\beta = 0.60$. }
   \label{fig:evolution}
\end{figure}

For an isolated halo, we assume spherical symmetry and use the following set of transport equations to describe the gravothermal evolution in the radial direction
\alg{
{}&\f{\partial}{{\partial r}}{M}=4\pi r^2 \rho, \quad \f{\partial}{\partial r} (\rho \nu^2) = -\frac{G M \rho }{r^2}, \\
{}&\f{\rho \nu^2}{\gamma-1} \left(\f{\partial}{\partial t}\right)_M \ln\f{\nu^2}{\rho^{\gamma-1}} = -\frac{1}{4\pi r^2} \f{\partial L}{\partial r} - C, \label{eq:4}
}
where $M(r,t)$ is the fluid mass enclosed within radius $r$ at a time $t$, $\rho(r,t)$ is the local density, $\nu(r,t)$ is the one-dimensional velocity dispersion, $L(r,t)$ is the luminosity, $C(r,t)$ is the volumetric bulk cooling rate, $G$ is the gravitational constant, and $(\partial_t)_M$ denotes the Lagrangian time derivative. The temperature is related to $\nu$ as $m\nu^2 = k_B T$, where $k_B$ is the Boltzmann constant. We assume the DM particle is monatomic and set the adiabatic index $\gamma=5/3$. The elastic and dissipative interactions are encoded in the conduction $\partial L/\partial r$ and the cooling term $C$, respectively. In this work, we assume both the elastic and inelastic cross sections are independent of the DM velocity.

DM elastic self-scattering allows radial heat conduction. This can be characterized by comparing the mean free path $\lambda=1/n\sigma$, where $n$ is the local number density and $\sigma$ is the cross section, to the scale height $H=\sqrt{\nu^2/4\pi G\rho}$. The ratio of $\lambda$ to $H$ is the Knudsen number, $Kn\equiv \lambda/H$, which indicates the importance of heat conduction induced by elastic scattering. We refer to regions with $Kn > 1$ ($Kn <1$) as long-mean-free-path (short-mean-free-path) regions. Note that $Kn\approx t_{r}/t_{\rm dy}$, where $t_r\approx \lambda/\nu$ is the local relaxation time for the elastic scattering and $t_{\rm dy}= H/\nu$ is the dynamical time of the halo. The luminosity $L$ is a function of the temperature gradient $L/4\pi r^2= -\kappa  {\partial T}/{\partial r}$, where the conductivity $\kappa = (\kappa_{\rm lmfp}^{-1}+\kappa_{\rm smfp}^{-1})^{-1}$ reduces to the conductivity of the long-mean-free-path ($\kappa_{\rm lmfp}$) and short-mean-free-path ($\kappa_{\rm smfp}$) regions in the appropriate limits, i.e., $\kappa_{\rm lmfp}  = (3\beta/2) n H^2 k_B/t_r \simeq 0.27\beta {n \nu^3 \sigma k_B}/({G m})$, and $\kappa_{\rm smfp} = (75\pi /256) n \lambda^2 k_B/t_r \simeq 2.1 {\nu k_B}/{\sigma}$~\cite{Balberg:2001qg,Balberg:2002ue,Koda:2011yb,Pollack:2014rja}. We determine the numerical factor $\beta$ in $\kappa_{\rm lmfp}$ by calibrating the fluid model model with $N$-body simulations. In this work, we have tested $\beta = 0.75,\;0.60,$ and $0.45$ for isolated~\cite{Koda:2011yb} and cosmological~\cite{Elbert:2014bma} $N$-body simulations with purely elastic DM self-interactions. Moreover, we have checked our fluid-model predictions ($\beta=0.60$) with recent dissipative SIDM $N$-body simulations~\cite{Choquette:2018lvq} and find good overall agreement; see the Supplemental Material for details.

Since we assume the energy released during the dissipative collision is not reabsorbed by DM particles in the halo, the cooling rate $C$ appears as a bulk term in~\eqref{4}, which can be written as a function of the model parameters,
\beq  \label{eq:bulk-cooling-rate}
C  = \vev{ \! \f{n E_{\rm loss}}{t'_r} \!} = \rho^2 \f{\sigma'}{m} \f{4\nu \nu_{\rm loss}^2}{\sqrt{\pi}} \left( 1 +\f{\nu_{\rm loss}^2}{\nu^2} \right) e^{-\f{\nu_{\rm loss}^2}{\nu^2}},
\eeq 
where $\nu_{\rm loss}\equiv\sqrt{E_{\rm loss}/m}$ is the ``velocity loss'' that parameterizes the energy loss per collision; $\sigma'$ is the cross section of the dissipative interaction and $t'_r \equiv 1/(n \sigma' v_\text{rel})$ is the relaxation time with respect to the relative velocity of the two incoming particles, $v_\text{rel}$;  we take the thermal average $\langle \cdot\rangle$ with respect to the Boltzmann distribution of $v_\text{rel}$ while restricting inelastic scattering to particles whose kinetic energy exceeds $E_{\rm loss}$, i.e., $v_\text{rel} \geq 2 \nu_\text{loss}$. This model of cooling captures the essential features of dissipative interactions. 

\begin{figure*}[t]
   \centering
   \includegraphics[width=0.96\textwidth]{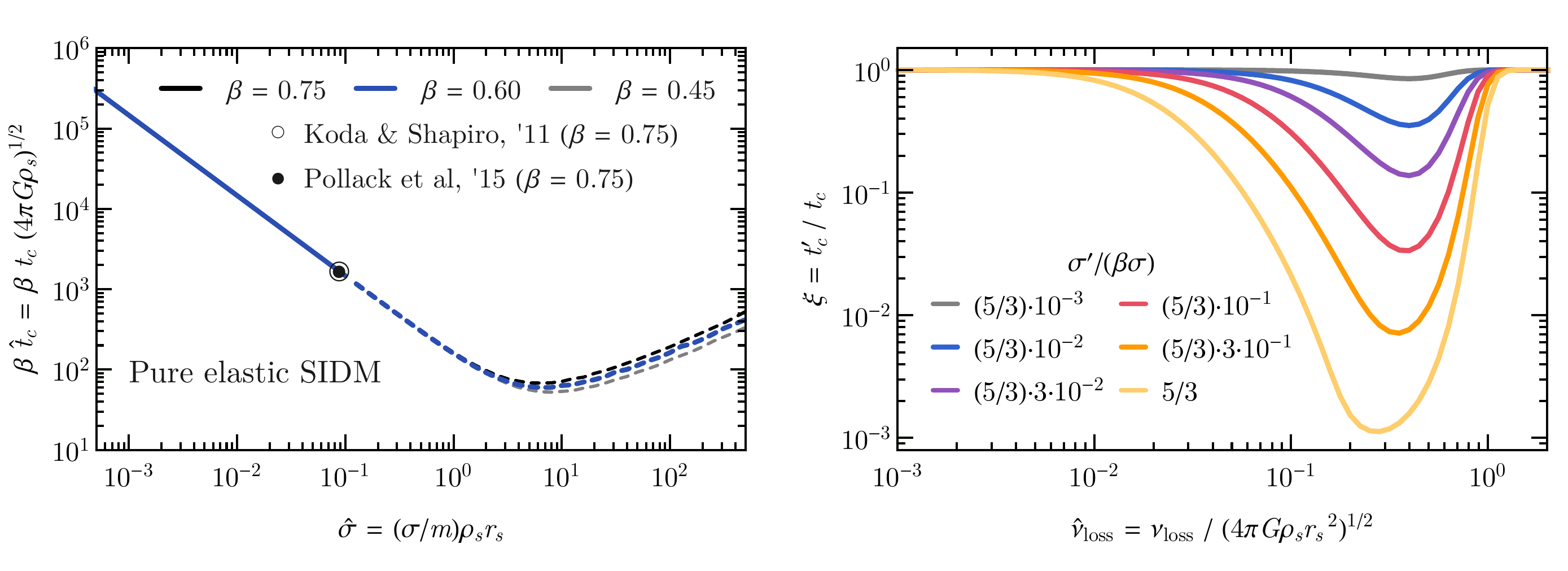}   
   \vspace{-0.4cm}
     \caption{{\bf Left:} Dimensionless collapse time $\hat t_c$ multiplied by $\beta$ as a function of $\hat \sigma$ when cooling is absent with $ \beta = 0.75$ (black), $0.60$ (blue) and $0.45$ (gray). We also show the results recasted from $N$-body simulations~\cite{Koda:2011yb} and fluid-model predictions~\cite{Pollack:2014rja}, where $\beta =0.75$ were used. In this work, we focus on $\hat \sigma \leq 10^{-1}$ (solid). {\bf Right:} Ratio of the collapse time, $\xi \equiv t'_c$/$ t_c$, as a function of $ \hat \nu_{\rm loss}$ for different values of $\sigma'/(\beta\sigma)$, where we take $\hat \sigma=10^{-2}$ with $\beta = 0.60$. For the range of $\hat\sigma=10^{-4}\textup{--}10^{-1}$ we have checked, $\sigma'/(\beta\sigma)$ well characterizes the $\xi\textup{--}\hat{\nu}_{\rm loss}$ relation. 
     }
   \label{fig:timescale00} 
\end{figure*}

We solve \eqref{4} with the boundary conditions at $t=0$ of $M=L=0$ for the inner boundary and $L=0$ for the outer boundary. We assume the initial halo mass distribution follows an NFW profile~\cite{Navarro:1996gj}, $\rho(r)=\rho_s r^3_s/r(r+r_s)^2$, where $\rho_s$ and $r_s$ are the scale density and radius, respectively. In our simulations, we reformulate~\eqref{4} in terms of a set of dimensionless variables based on $r_s$ and $\rho_s$ and follow the numerical procedure in Refs.~\cite{Balberg:2002ue,Pollack:2014rja}; see the Supplemental Material.

\stepcounter{sec}
{\emph {Gravothermal Evolution.}---\;}To illustrate the effect of the dissipative interactions, we consider a dwarf halo with mass $8\times10^{10}M_\odot$ and characteristic halo parameters $r_s=6.5~{\rm kpc}$ and $\rho_s=1.28\times10^7~M_\odot/{\rm kpc^3}$. Reference~\cite{Kamada:2016euw} took this NFW halo as an outer boundary condition to find the SIDM fit to the galactic rotation curve of LSB F583-1, which exhibits a cored density profile. We take $\sigma/m=3~{\rm cm^2/g}$ as in Ref.~\cite{Kamada:2016euw} and consider $\sigma'=0$ as well as $\sigma'=\sigma$ and $\nu_{\rm loss}=13~{\rm km/s}$, corresponding to the benchmark case 2 as we will discuss below.

\figref{evolution} shows the density vs radius over time with (red) and without (blue) bulk cooling. Each curve is labeled with a $Kn$ value for the innermost simulated shell. From the density evolution, we see the process can be divided into three stages: (i) Core expansion. Heat conduction is inwards ($L<0$) and ${Kn} \gg 1$. The halo evolves quickly to a quasi-isothermal state. (ii) Self-similar collapse. Heat is conducted outwards ($L>0$) and $Kn$ slowly decreases. The self-similar collapse results in a cuspy density profile and with log-slope of approximately $-2$, a characteristic feature if the cooling is absent or mild. (iii) Post-self-similar collapse. Here $Kn <1$ at the center and the inner density suddenly begins to increase dramatically. In~\figref{evolution}, the symbol diamond denotes the $1\to2$ transition, when the innermost shell is at its least dense and its luminosity vanishes; the symbol star denotes the $2\to3$ transition, when $Kn=1$.

For concreteness, we define a collapse time as the time at which $Kn=0.1$ for the innermost shell, and we denote the collapse time with (without) inelastic cooling as $t'_c$ ($t_c$). 
 Since the evolution of the third stage is very fast, $t'_c$ and $t_c$ are largely determined by the time of the first two stages. The most important effect of the dissipative interaction is to significantly speed up the collapse time, $t_c' < t_c$. For LSB F583-1 with the model parameters chosen in Fig.~\ref{fig:evolution}, the collapse time with cooling is shortened by about a factor of $20$, resulting in $t_c'\approx 8.5~{\rm Gyr}$. This amount of cooling is disfavored, because the final density profile is too steep to be consistent with the observed profile of LSB F583-1~\cite{Kamada:2016euw}.   
 
We perform a suite of simulations, varying the model parameters within the following range of values in dimensionless units: $\hat \sigma\equiv(\sigma/m)\rho_s r_s=10^{-4}\textup{--}10^{3}$, $\sigma'/\sigma=10^{-3}\textup{--}1$, and $\hat \nu_{\rm loss}\equiv \nu_{\rm loss}/(4\pi G\rho_s r^2_s)^{1/2}=0\textup{--}5$ with evenly log-spaced steps, and $\beta = 0.45,\;0.60,$ and $0.75$. In~\figref{timescale00} (left), we show results for the halo evolution with pure elastic self-scattering and no cooling.  For $\hat \sigma \lesssim1$, there is a simple scaling relation between 
$\beta \hat t_c \equiv \beta {(4\pi G \rho_s)^{1/2}} \, t_c$ and $\hat\sigma$, namely $\beta \hat t_c\approx 150/\hat \sigma$, which can be expressed as 
\begin{equation}
t_c\approx \frac{150}{\beta}\frac{1}{r_s\rho_s\sigma/m} \frac1{\sqrt{4\pi G\rho_s}}\,.
\label{eq:tcnocooling}
\end{equation}
In this regime, a large $\hat\sigma$ speeds up the thermal evolution of the halo and shortens the collapse timescale. However, as $\hat\sigma\gtrsim1$, the inverse proportionality is lost because the mean free path is too short and heat conduction is actually suppressed~\cite{Gnedin:2001gd,Agrawal:2016quu}. Below, when setting constraints on dissipative DM, we restrict to $\hat \sigma \leq 0.1$, along with $\hat \sigma' \leq\hat\sigma\leq 0.1$, so that the mean free path is larger than the scale height for the halos we consider. Thus, the initial halo is in the optically thin regime and the cooling effect is mild. For the parameters shown in \figref{evolution}, $\hat \sigma = 0.1$ corresponds to $\sigma/m=0.1/r_s \rho_s=5.8 \cmg$, so the  choice of $\sigma/m=3\cmg$ satisfies the condition. Note we can recast the scaling relation in Eq.~(\ref{eq:tcnocooling}) as $t_c\propto r^{-1}_s\rho^{-3/2}_s\propto M^{-1/3}_{200}c^{-7/2}_{200}$, where $M_{200}$ and $c_{200}$ are the halo mass and concentration~\cite{Navarro:1995iw}, respectively. Thus, $t_c$ is extremely sensitive to $c_{200}$, which may have important implications for understanding dwarf galaxies in the Milky Way~\cite{Nishikawa:2019lsc,Kaplinghat:2019svz,Sameie:2019zfo,Kahlhoefer:2019oyt}. 

Our results are in good agreement with Ref.~\cite{Pollack:2014rja}, where $\hat \sigma = 0.088$ and $\beta = 0.75$ were chosen. To compare with cosmological $N$-body simulations of dwarf halos in Ref.~\cite{Elbert:2014bma}, we take the Pippin halo parameters, $r_s=2.7~{\rm kpc}$ and $\rho_s=1.73\times10^{7}~M_\odot/{\rm kpc^3}$, and apply Eq.~(\ref{eq:tcnocooling}). The estimated core-collapse time is $t_c\approx 80~{\rm Gyr}$ for $\sigma/m =10~\cmg$ and $t_c\approx 16~{\rm Gyr}$ for $\sigma/m =50~\cmg$ for $\beta=0.60$, consistent with the absence of core collapse and the presence of a mild collapse, respectively, observed in the simulations. We also find that a calibration with $\beta=0.45$ yields a better agreement with the cosmological simulations.

\figref{timescale00} (right) shows the reduction of the collapse time, $\xi\equiv t'_c/t_c$, from dissipative interactions. We find that the $\xi\textup{--}{\hat\nu}_{\rm loss}$ relation is well characterized by $\sigma' /(\beta\sigma)$ for the test values, i.e., $\hat \sigma=10^{-4}\textup{--}10^{-1}$ and $\beta=0.45\textup{--}0.75$. Overall, the maximal reduction is achieved when $\hat\nu_{\rm loss} \approx0.3$ for a wide range of $\sigma' /(\beta\sigma)$. 

The origin of this scale can be understood as the following: as the evolution starts, the cold inner halo ($r< r_s$) quickly thermalizes with the maximum velocity-dispersion of the initial NFW profile, which is about $\nu \sim 0.3\ (4\pi G \rho_s r_s^2)^{1/2}$ at $r=r_s$ [or $\hat \nu \equiv \nu/(4\pi G \rho_s r_s^2)^{1/2} \sim 0.3$], and stays near that value for most of the halo's evolution. For $\hat \nu_{\rm loss} \lesssim 0.3$, the energy loss is small per collision, while for $\hat \nu_{\rm loss} \gtrsim 0.3$, inelastic scattering can only occur among particles on the high-velocity tail or very late in the halo evolution (stage 3).

The collapse time in the presence of cooling is then $t'_c =\xi[\sigma'/(\beta\sigma),\hat \nu_{\rm loss}] t_c$, where $t_c$ is given by \eqref{tcnocooling} and $\xi$ can be read from Fig.~\ref{fig:timescale00} (right). For $\hat \nu_{\rm loss} < 0.2$, we find an approximate formula $\xi \approx \exp\left[ - {\nu_{\rm loss} \sqrt{\sigma'/(\beta \sigma)} }/{0.035 (4\pi G\rho_s r^2_s)^{1/2}}  \right]$. The collapse time can be reduced as much as a factor of $10^3$, indicating that dissipative scattering can be important for the evolution of the SIDM halo. Compared to $t_c$, $t_c'$ is also sensitive to $\nu_{\rm loss} = \hat \nu_{\rm loss} (4\pi G \rho_s r^2_s)^{1/2}\propto c^{1/2}_{200}M^{1/3}_{200}$.

\stepcounter{sec}
{\emph {Astrophysical Implications.}---\;}\label{sec:astro}For many dwarf and LSB galaxies, DM dominates the dynamics, and the stars and gas particles trace the gravitational potential well of the halo. Reference~\cite{Kamada:2016euw} analY.Z.ed the rotation curve data of $30$ spiral galaxies and found that they can be fitted with an SIDM model with {\it elastic} self-scattering cross section $\sigma/m=3~{\rm cm^2/g}$. Among them, $18$ of the galaxies have low baryon content and also exhibit a constant density core with no evidence of gravothermal collapse. We use this sample to constrain the dissipation parameters $\sigma'$ and $\nu_{\rm loss}$ by demanding $t_c'>10~{\rm Gyr}$.

\begin{figure}[t]
   \centering
   \includegraphics[width=0.44\textwidth]{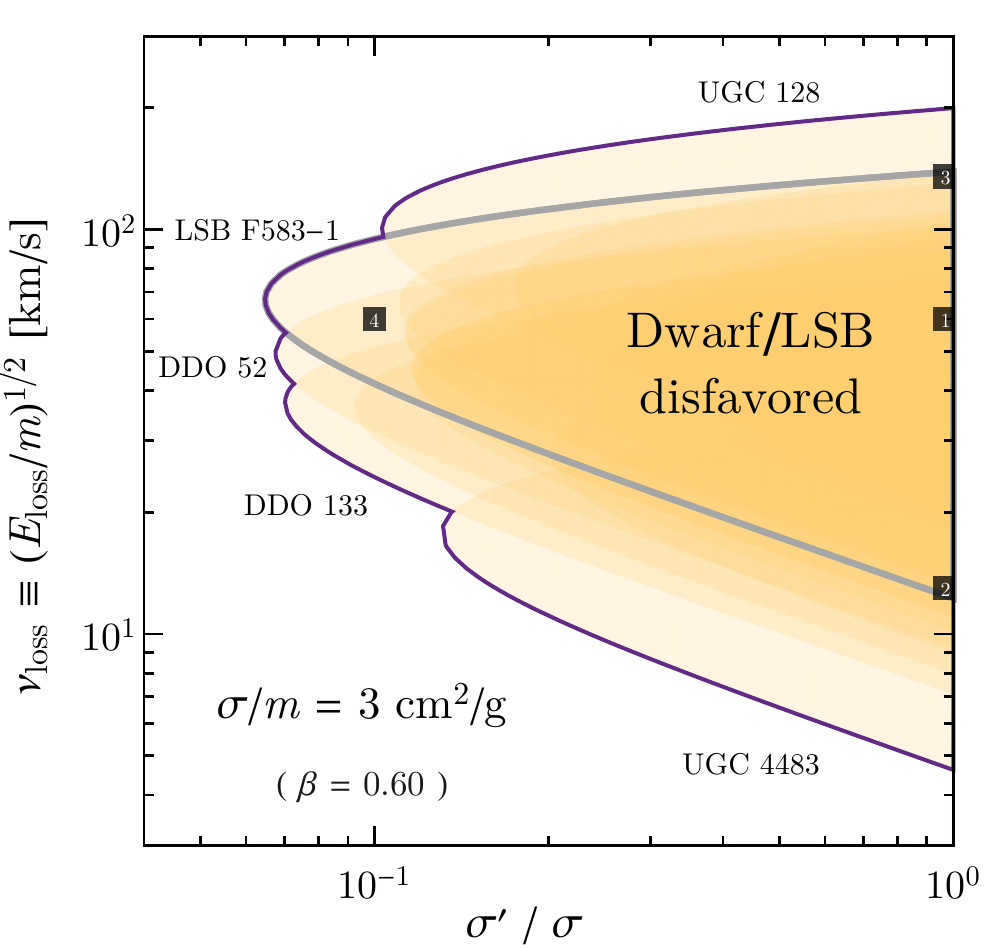} 
   \vspace{-0.2cm}
   \caption{Constraints on the dissipative parameters from the absence of core collapse in individual dwarf galaxies (yellow) within $10~{\rm Gyr}$ and its overall boundary for the sample (purple). We take the fitted halo parameters of the galaxies from Ref.~\cite{Kamada:2016euw}, also listed in the Supplemental Material and labeled for the outer ones, and $\beta=0.60$ in our fluid simulations. See text for detailed discussion on LSB F583-1 (gray) and the four benchmark cases (marked with numbers). We focus on the $\sigma'/\sigma \leq 1$ region, where the fluid model is applicable.}
   \label{fig:constraints}
\end{figure}

\figref{constraints} shows regions (shaded) where core collapse occurs in less than $10~{\rm Gyr}$ for individual galaxies, taking halo parameters $r_s$ and $\rho_s$ from~\cite{Kamada:2016euw} as input. In these regions, the inner density profiles of the associated halos at $10~{\rm Gyr}$ are much steeper than inferred from the stellar kinematics~\cite{Kamada:2016euw}; see Supplemental Material for more details on determining the exclusion limits. In solid  purple, we show the boundary from all galaxies imposing the constraint $t'_c<10\,{\rm~Gyr}$ with calibration parameter $\beta = 0.60$; roughly the region with $0.1 \lesssim \sigma'/\sigma \lesssim 1$ and $10\,{\rm km}/{\rm s}\lesssim \nu_{\rm loss}\lesssim 100 \,{\rm km}/{\rm s}$ is disfavored. We have checked the constraints are insensitive to the $\beta$ values considered in the work; see Supplemental Material for additional results with $\beta=0.45$. 

We explicitly demonstrate how the results in Fig.~\ref{fig:timescale00} can be used to derive the constraints in Fig.~\ref{fig:constraints}. Take LSB F583-1 as an example and focus on four benchmark points shown with small squares in Fig.~\ref{fig:constraints}. For pure elastic DM self-interactions with $\sigma/m=3\cmg$ and $\beta=0.60$, $t_c\approx1.7\times 10^2~{\rm Gyr}$ from~\eqref{tcnocooling}, much longer than the age of the Universe. Taking $\sigma'/\sigma=1$ and $\nu_{\rm loss}=60~{\rm km/s}$, $\hat \nu_{\rm loss}\approx0.3$, so that $\xi\approx10^{-3}$ from Fig.~\ref{fig:timescale00}, resulting  in  a  much shorter collapse  time, $t'_c=\xi t_c\approx0.2~{\rm Gyr}$. Keeping $\sigma'/\sigma=1$, and taking $\nu_{\rm loss}=13~{\rm km/s}$ ($135~{\rm km/s}$), so that $\hat \nu_{\rm loss}\approx0.075$  ($0.78$), we find $\xi\approx 0.049\,(0.043)$ leads to $t_c'\approx 8.5  (7.6)\, {\rm Gyr}$, which is disfavored. Finally, for $\sigma'/\sigma=0.1$ and $\nu_{\rm loss}=60~{\rm km/s}$, we find $\hat \nu_{\rm loss}\approx0.35$ and $\xi\approx 0.035$. This gives $t_c'\approx 6.3~{\rm Gyr}$, which again is disfavored.

 As an application, we consider the atomic DM model with hyperfine splitting transitions~\cite{Boddy:2016bbu}. If the dark proton is the much heavier than the dark electron, $\nu_\text{loss} =\sqrt{2E_{\rm hf}/m_{p}}= \sqrt{3/4}E_{\rm hf}/E_0\approx27~{\rm km/s}$, where $m_{p}$ is the dark proton mass, and $E_{\rm hf}/E_0=10^{-4}$ is the ratio of the hyperfine splitting to the binding energy. In this case, $\sigma/m_{p}\gtrsim1~{\rm cm^2/g}$ on dwarf scales, and $\sigma'/\sigma$ can be in the range of $0.1\textup{--}1$ for the dark structure constant $\sim0.02\textup{--}0.08$ and $m_{p}\sim10\textup{--}60~{\rm GeV}$~\cite{Boddy:2016bbu}. Thus, this model is subject to the constraints shown in Fig.~\ref{fig:constraints}. In addition, our results also put a lower limit on the threshold velocity, $\gtrsim100~{\rm km/s}$, for dissipative SIDM models proposed to explain the formation of supermassive black holes~\cite{Choquette:2018lvq}.
 
 Our constraints shown in Fig.~\ref{fig:constraints} are based on $\sigma/m=3~{\rm cm^2/g}$. For models with different $\sigma/m$ values, we can apply the same procedure to derive corresponding constraints. Since the SIDM fits vary mildly for $\sigma/m=1\textup{--}10~{\rm cm^2/g}$~\cite{Ren:2018jpt}, we expect our method and results to have broad applications. In addition, we note that Ref.~\cite{Kamada:2016euw} has imposed strong constraints on the halo parameters $r_s$ and $\rho_s$ ($M_{200}$ and $c_{200}$) from cosmological simulations~\cite{Dutton:2014xda} and the sample of the $18$ galaxies we consider covers a wide range of halo concentration. Thus, our results are robust in the cosmological context.

\stepcounter{sec}

{\emph {Conclusions.}---\;}We have studied the gravothermal evolution of DM halos in the presence of dissipative DM self-interactions. After introducing a simple but well-motivated model to capture the cooling effect, we performed numerical simulations and obtained numerical templates between the core-collapse time and the model parameters, which can be easily adapted for specific particle physics realizations of dissipative DM. Utilizing the density cores inferred from the dwarf galaxies, we put strong constraints on the dissipation parameters. Our results have been overall confirmed by recent $N$-body simulations with dissipative SIDM~\cite{Choquette:2018lvq}. It is of interest to generalize our analysis to include velocity-dependent cross sections, which we leave for future work. Our formalism can be extended also to other scenarios, e.g., those proposed in Refs.~\cite{McDermott:2017vyk,Gresham:2018anj,Vogelsberger:2018bok,Chu:2018nki}, where DM particles are heated from energy release due to dark-sector interactions, which could further increase the core size of the halo and lower its central density.\\

We thank Jun Koda and Paul Shapiro for providing many insights on the fluid simulations as well as additional information on their $N$-body simulations. We also thank Jason Pollack for helpful discussions on the fluid simulations, as well as Prateek Agrawal, Joseph Bramante, Francis-Yan Cyr-Racine, Manoj Kaplinghat, Julio Navarro, Annika Peter, Ben Safdi, Martin Schmaltz, Neelima Sehgal, Scott Tremaine, and Sean Tulin for useful discussions.  The authors thank the KITP at UCSB for hospitality, where their research was supported by the National Science Foundation under Grant No. NSF PHY-1748958.  H.B.Y. acknowledges T.~D.~Lee Institute, Shanghai, and Y.Z. thanks the Aspen Center for Physics, for hospitality during the completion of this work. R.E. acknowledges support from U.S.~Department of Energy under Grant No.~DE-SC0017938.  S.D.M. was supported by the Fermi Research Alliance, LLC under Contract No. DE-AC02-07CH11359 with the U.S. Department of Energy, Office of Science, Office of High Energy Physics. H.B.Y. acknowledges support from U.S.~Department of Energy under Grant No. DE-SC0008541 and a UCR Regents' Faculty Development Award.  Y.Z. acknowledges support from U.S.~Department of Energy under Grant No. DE-SC0015845. The U.S. Government retains and the publisher, by accepting the article for publication, acknowledges that the U.S. Government retains a non-exclusive, paid-up, irrevocable, world-wide license to publish or reproduce the published form of this manuscript, or allow others to do so, for U.S. Government purposes.

\bibliography{sidm.bib}
\bibliographystyle{h-physrev.bst}

\newpage

\onecolumngrid

\begin{center}
\textbf{\large Constraining Dissipative Dark Matter Self-Interactions} \\ 
\vspace{0.05in}
{ \it \large Supplemental Material}\\ 
\vspace{0.1in}
{}
\vspace{0.05in}
{Rouven Essig$^{1}$, Samuel D. McDermott$^{2}$, Hai-Bo Yu$^{3}$, Yi-Ming Zhong$^{4}$}
\end{center}
\centerline{{\it  $^{1}$C.N. Yang Institute for Theoretical Physics, Stony Brook University, Stony Brook, NY 11794}}
\centerline{{\it  $^{2}$Fermi National Accelerator Laboratory, Center for Particle Astrophysics, Batavia, IL 92376}}
\centerline{{\it  $^{3}$Department of Physics and Astronomy, University of California, Riverside, CA 92521}}
\centerline{{\it  $^{4}$Physics Department, Boston University, Boston, MA 02215}}
\vspace{0.2in}

\twocolumngrid

We provide additional information and results, including the derivation of the cooling rate (A), the procedure for numerical simulations (B), the condition for the mild-cooling regime (C), calibration with isolated and cosmological $N$-body simulations (D), the halo parameters of $18$ dwarf and LSB galaxies taken from~\cite{Kamada:2016euw}, which were used to derive the bounds shown in Fig.~3 in the main text  (E), snapshots of density profile for the benchmark points of LSB F583-1 (F), and additional details on our limit-setting procedure (G). 

\setcounter{sec}{1}
\section{\Alph{sec}.~The Cooling Rate}
\label{sec:coolingrate}
The volumetric cooling rate $C$ is defined as the energy loss per unit volume per unit time, which can be calculated as 
\beq
 C= \vev{\f{n E_{\rm loss}}{t'_r} }  = \vev{{\rho^2 (\sigma'/m) v_{\rm rel}\nu_{\rm loss}^2} }, 
 \label{eq:C}
\eeq
where $\vev{\ldots}$ denotes the thermal average with respect to the relative velocity of the two incoming particles $v_{\rm rel}\equiv |\vec v_{\rm rel}|\equiv|\vec v_1-\vec v_2|$. $t'_r \equiv 1/(n \sigma' v_\text{rel})$ is the relaxation time for the dissipative interaction before thermal average. Assuming the dark matter velocity follows the Boltzmann distribution 
\beq
f(\vec v) = \f{1}{(2\pi)^{3/2} \nu^3} e^{-\f{v^2}{2 \nu^2}},
\eeq
the thermal average over the relative velocity for quantity $X$ is given by
\begin{align}
\vev{X} 
={}&\f{1}{2 \pi^{1/2} \nu^3}  \int^\infty_0 d v_{\rm rel} v_{\rm rel}^2 e^{-\f{v_{\rm rel}^2}{4\nu^2}} X.
\end{align}
In computing the cooling rate, we impose a lower cut-off on $v_{\rm rel}$ to take into account the fact that the energy loss during an inelastic collision should be smaller than the total kinetic energy of the two colliding particles in the center-of-mass frame,
\beq
E_{\rm loss} \leq \f{1}{2}\mu v_{\rm rel}^2 = \f{1}{4}m v_{\rm rel}^2 \Rightarrow v_{\rm rel} \geq  2 \nu_{\rm loss} 
\eeq
where $\mu = m/2$ is the reduced mass. Thus, the cooling term can be written as
\beq
C= \frac{\rho^2\nu_{\rm loss}^2}{2\pi^{1/2} \nu^3} \int_{2 \nu_{\rm loss}}^\infty d v_{\rm rel}  v_{\rm rel}^2 e^{- \f{v_{\rm rel}^2}{4\nu^2}} { (\sigma'/m) v_{\rm rel}}.
\eeq
If $\sigma'$ is velocity-independent, we have
\beq
C =\vev{\f{n E_{\rm loss}}{t'_r} }_{T\geq E_{\rm loss}} \!\!\!= \f{4}{\sqrt{\pi}}  \rho^2 \f{\sigma'}{m} \nu \nu_{\rm loss}^2 \left(1 +\f{\nu_{\rm loss}^2}{\nu^2}\right) e^{-\f{\nu_{\rm loss}^2}{\nu^2}}.
\label{eq:fullc}
\eeq
The cooling effect is small when $\nu \ll \nu_{\rm loss}$, and it becomes significant when  $\nu > \nu_{\rm loss}$.

\stepcounter{sec}
\section{\Alph{sec}.~Numerical Simulations}
\label{sec:numerical}

In performing our numerical simulations, we have used dimensionless variables by taking the ratio of a physical quantity ($x$) to its corresponding fiducial value, i.e., $\hat x \equiv x/x_0$, and then rewrote Eq.~(1) in the main text into dimensionless form. The fiducial quantities are built from the halo parameters $\rho_s$ and $r_s$, as shown in~\tabref{fiducial}. 
\begin{table}[h]
   \centering
      \topcaption{Fiducial quantities used in our numerical simulations.}
   \begin{tabular}{@{} |cc| @{}} 
      \hline
       $M_0 = 4\pi  \rho_s r_s^3$ &  $(\sigma/m)_0 = ({r_s \rho_s})^{-1}$     \\
       $\nu_0 = (4\pi G \rho_s)^{1/2}  r_s$  & $L_0 = (4\pi)^{5/2} G^{3/2} \rho_s^{5/2} r_s^5\label{eq:L0}$  \\
$t_{0} =(4\pi G \rho_s)^{-1/2}$ & $C_0 =  (4\pi G)^{3/2} \rho_s^{5/2} r_s^2\label{eq:C0}$   \\ 
      \hline
   \end{tabular}
   \label{tab:fiducial}
\end{table}

In the fluid model, a self-gravitating halo is assumed to be in hydrostatic equilibrium. The conservation of mass, momentum, and energy resembles the zeroth, first, and the second moments of the Boltzmann equation. Under the assumption that the velocity dispersion is spherically symmetric, they form a closed set of equations that truncates the Bogoliubov hierarchy. We segmented the halo into $N=150$ evenly log-spaced concentric shells in radius $\{\hat r_1, \hat r_2, \cdots, \hat r_N\}$ with $\hat r_1 = 10^{-2}$ and $\hat r_N = 10^3$. Following the treatment in Ref.~\cite{Pollack:2014rja}, we take values of the extensive quantities ($\hat M_i$, $\hat L_i$) at the radius of $i$-th sphere and those of the intensive ones ($\hat \rho_i, \hat \nu_i, \hat C_i$) as the average between $i$-th and $(i-1)$-th spheres. The algorithm uses Lagrangian zones: the radii $\hat r_i$ are allowed to change, but the mass in each shell is fixed. 

Each step of the evolution is separated into two stages: \emph{thermal energy} is exchanged by conduction and/or cooling, after which  \emph{hydrostatic relaxation} brings the system back to equilibrium. We assume that the specific entropy, $s\propto \ln (\nu^3/\rho)$, is conserved during hydrostatic relaxation. The workflow is as follows:
\begin{enumerate}
\item Compute the luminosity and the cooling profiles $\hat L_i$ and $\hat C_i$ based on profile input $\hat r_i$, $\hat \rho_i$, and $\hat \nu_i$, and particle physics input $\hat \sigma$, $\hat \sigma'$, and $\hat \nu_{\rm loss}$.
\item Allow a small passage of time $\Delta \hat t$ and compute the specific energy change $\Delta \hat u_i$ ($\hat u_i \equiv 3\hat \nu_i^2/2$) by conduction and cooling, assuming fixed  density. Eq.~(1) in the main text gives 
\beq
\f{\Delta \hat u_i}{\Delta \hat t} =  - \left(\frac{\partial \hat L}{\partial \hat M}\right)_i -\f{\hat C_i}{\hat \rho_i},
\label{eq:5}
\eeq
and we update $\hat u_i$ by $\Delta \hat u_i$. The time step $\Delta \hat t$ is sufficiently small  ($|\Delta \hat u_i/\hat u_i| < 10^{-3}$) 
such that the linear approximations used in step 3 below are valid.

\item Upon updating $\hat u_i$, the $i$-th shell is no longer virialized. To return to hydrostatic equilibrium, we perturb $\hat r_i$, $\hat \rho_i$, and $\hat \nu_i$ while keeping the mass $\hat M_i$ and specific entropy $\hat s_i = \ln (\hat \nu_i^3/\hat \rho_i)$ of the shell fixed.  We treat mass conservation, hydrostatic equilibrium relations, and energy conservation, shown in Eq.~(1) in the main text, at the linear order and solve them for all shells simultaneously. For numerical accuracy, we iteratively perform the perturbation 10 times until hydrostatic equilibrium is established everywhere.  

\item Re-establishing hydrostatic equilibrium gives new values for $\hat r_i$, $\hat \rho_i$, and $\hat \nu_i$. We return to step 1 and update the luminosity $\hat L_i$ and cooling profile $\hat C_i$.

\item Track the $Kn$ for the innermost shell. The evolution is terminated when $Kn$ drops below 0.1 (stage 3).

\end{enumerate}
The above procedure is coded in \texttt{C++} with the \texttt{eigen 3.2.8} library for linear algebra~\cite{eigenweb}. 

\stepcounter{sec}
\section{\Alph{sec}.~Strong Cooling vs. Mild Cooling}
\label{sec:strongcooling}
Cooling is strong if the kinetic energy gain of the infalling particles can be efficiently removed from the halo on a time scale smaller than the free-fall time, $t_{\rm ff} = \sqrt{3\pi/(32 G {\rho})}$, which is close to $t_{\rm dy}$. This is similar to the ``isothermal collapse" or ``free-fall collapse" of the evolution of protostars (see e.g.~\cite{mo2010galaxy, binney2011galactic}). A halo under isothermal collapse can fragment into multiple dark clumps and may also lead to the formation of a dark disk or a dark bulge. The strong cooling condition is set by $t_{\rm cool} \lesssim t_{\rm ff}$, where the cooling timescale is $t_{\rm cool} = (3/2) {\rho}{\nu}^2/C$. In the strong cooling regime, the inner part of a dwarf-size halo will collapse and fragment within a time of $t_{\rm ff} \sim 0.1~{\rm Gyr}$ for typical densities of $\rho_s = 10^6-10^7 \mkpc$.  Since $t_{\rm ff} \ll 10~{\rm Gyr}$, the corresponding  parameters for strong cooling can be excluded by dwarf galaxy observations. 

By definition, in the strong cooling regime the timescale of the thermal energy change ($t_{\rm cool}$) is smaller than that of hydrostatic relaxation ($t_{\rm dy}$). This invalidates the assumptions in the fluid simulations that the system stays in hydrostatic equilibrium. Hence, we should only trust our simulations for parameters that provide $t_{\rm cool} \gg t_{\rm dy}$.
As we show now, mild cooling indeed describes much of our parameter space.

The ratio between $t_{\rm cool}$ and $t_{\rm ff}$ is given by
\begin{align}
\frac{t_{\rm cool}}{t_{\rm ff}} = \frac{ 3{\rho} {\nu}^2 }{2C} \sqrt{\f{32 G {\rho}}{3\pi}} = \frac{\sqrt{\frac{3}{8\pi}}(\hat \sigma/\hat \sigma') {Kn}}{ \f{\hat \nu_{\rm loss}^2}{\hat {\bar \nu}^2} \left(1 +\f{\hat \nu_{\rm loss}^2}{\hat {\bar \nu}^2}\right) e^{-\f{\hat \nu_{\rm loss}^2}{\hat {\bar \nu}^2}}}.
\label{eq:tcooltdy}
\end{align}
We take the initial density profile to be given by the NFW profile and $\hat \sigma' \leq 0.1$. For $\hat \sigma' =0.1$, the ratio $t_{\rm cool}/t_{\rm ff}$ at the inner most region ($r = 0.01 r_s$) is $\geq 3$ for all possible $\hat \nu_{\rm loss}$ and $\geq 10$ for either $\hat \nu_{\rm loss}\leq 0.07$ or $\geq 0.3$. At larger radii ($r\geq 0.01 r_s$), $Kn$ monotonically increases while $\nu$ monotonically decreases. Both factors contribute to a fast growth of 
$t_{\rm cool}/t_{\rm ff}$. Reducing $\hat \sigma'$ increases $t_{\rm cool}/t_{\rm ff}$. Consequently, this extends the range of $\hat \nu_{\rm loss}$ that permits $t_{\rm cool}/t_{\rm ff}\gg 1$ and for which our analysis is valid. We explicitly checked that the parameter space for the $18$ galaxies that saturate $t_c = 10~\rm{Gyr}$  with $\beta=0.60$ and $0.45$ all have $t_{\rm cool}/t_{\rm ff} > 30 \gg 1$, i.e., the halo collapse begins when the cooling is still mild.

As the evolution proceeds and enters stage 3, $Kn$ for the innermost region drops below 1. This means that at this point of the evolution, $t_{\rm cool}/t_{\rm ff}\,\propto Kn$ for the inner part of the halo also drops below 1, and the inner core may begin to fragment as it isothermally collapses. 

Nevertheless, given that the pre-factor $\sqrt{3/8\pi}(\hat \sigma/\hat \sigma')[ (\nu_{\rm loss}^2/\nu^2)(1+\nu_{\rm loss}^2/\nu^2)\exp(-\nu_{\rm loss}^2/\nu^2)]^{-1}$ is greater than 1 for $Kn > 2.4$, the isothermal collapse is unlikely to happen before the very end of stage 2. Thus, it does not significantly affect our fluid-model-based collapse time estimation, nor does it affect the conclusion of a cuspy inner density profile. On the other hand, different techniques are necessary to understand the ultimate fate of the matter in this region and whether or not gravitational collapse to a massive black hole is possible. 

\stepcounter{sec}
\section{\Alph{sec}.~Calibration and Cross-check with $N$-body Simulations}
\label{sec:calibration}
Unlike the conductivity in the small mean-free-path (smfp) region that is well-described by kinetic theory~\cite{pitaevskii2013course}, the conductivity in the long mean-free-path (lmfp) region, $\kappa_{\rm lmfp}$,  cannot be fully determined from first principles. The existence of the free coefficient $\beta$ in the expression of $\kappa_{\rm lmfp}$ reflects this fact (see discussion below Eq.~(2) in the main text). Its value can be fixed by comparing the evolution of the inner density profile from the fluid model to that of the $N$-body simulations.  This is 
possible, since $\kappa_{\rm lmfp} \propto \beta$ determines the collapse time of stage 2, which takes the bulk of the entire evolution time.

\begin{figure}[t!]
   \centering
   \includegraphics[width=0.4\textwidth]{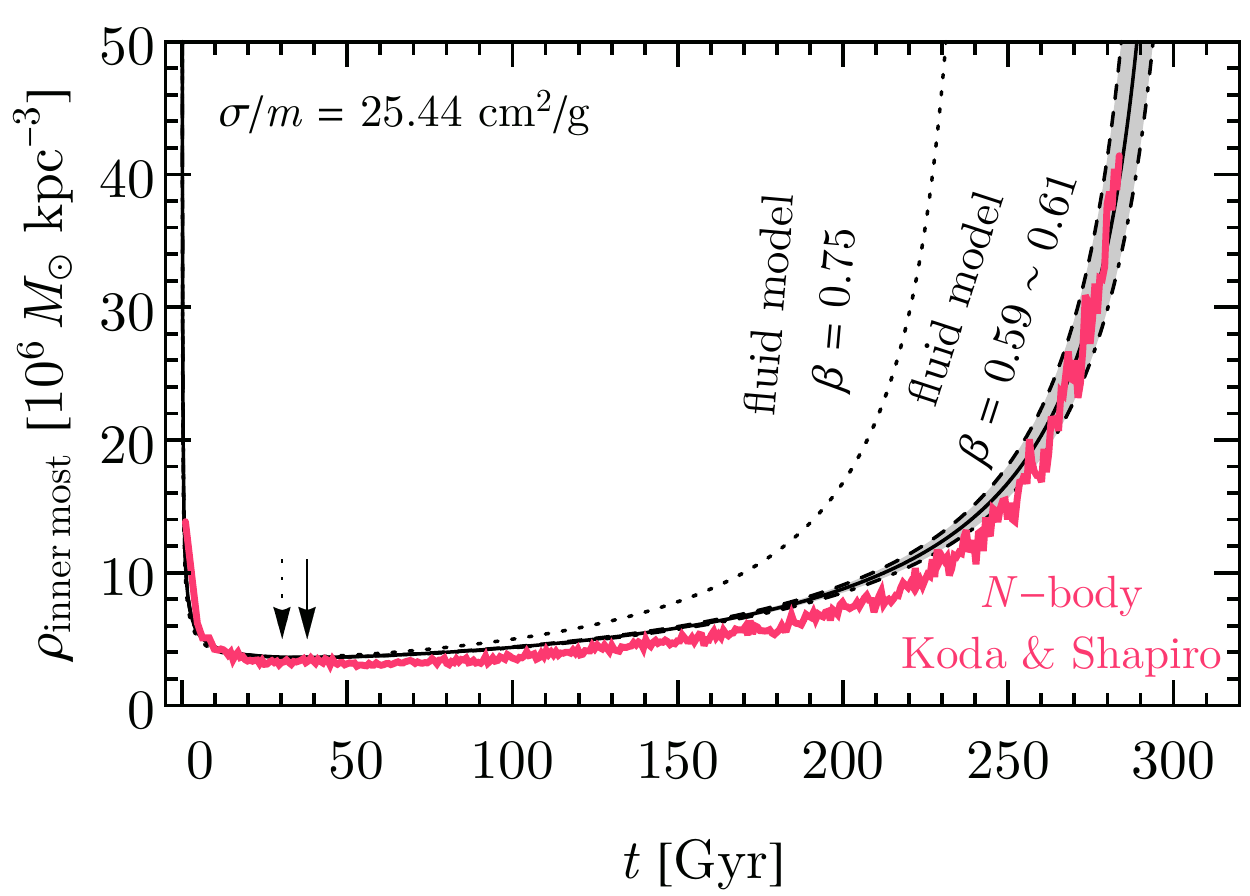}    \caption{Calibrating coefficient $\beta$  using the central density as a function of the evolution time. Results from an isolated $N$-body simulations are shown as the red line~\cite{Koda:2011yb,private} . Results from fluid simulations with various $\beta$ are shown as black lines (dotted: $\beta=0.75$, dashed: $\beta =0.59$, solid: $\beta = 0.60$, dot-dashed: $\beta=0.61$). The solid (dashed) downward arrow indicates the moment for stage $1 \to 2$ transition for $\beta = 0.59-0.61$ ($\beta = 0.75$).}
   \label{fig:calibrationkoda}
\end{figure}

Various calibrations have been presented in the literature. Ref.~\cite{Balberg:2002ue, Koda:2011yb} used $N$-body simulations of an isolated halo with an initial profile given by a self-similar core-like profile (a.k.a., the Balberg-Shapiro-Inagaki  profile) and find $\beta\simeq 0.75$. However, Ref.~\cite{Koda:2011yb} also showed that when applying $\beta=0.75$ to a halo with an initial profile given by the NFW profile, the collapse time inferred from a fluid model is about $\sim20\%$ shorter than that from the $N$-body simulations.  We re-examine this calibration by using our own analysis of the fluid model, which we compare to the $N$-body simulation data of Ref.~\cite{Koda:2011yb}. The parameters for the simulated halo are $\rho_s = 1.49\times 10^6 \mkpc$ and $r_s = 11.1 {\rm\,kpc}$, with an elastic cross section $\sigma/m = 25.44 \cmg$~\cite{private}. We find that $0.59\leq \beta \leq 0.61$ agrees well all three stages of the evolution, see \figref{calibrationkoda}. Note that the difference between $\beta = (0.59, 0.60, 0.61)$ only becomes evident near the end of stage 2. 

\begin{figure*}[t]
   \centering
   \includegraphics[width=0.4\textwidth]{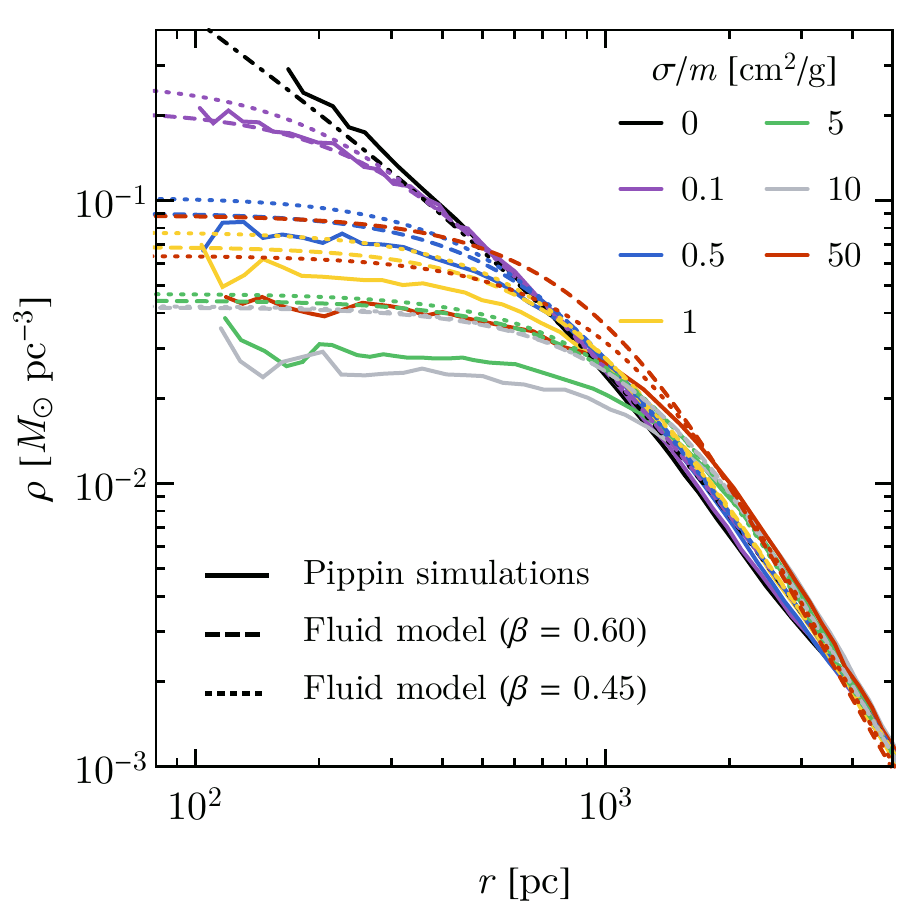}~~   \includegraphics[width=0.4\textwidth]{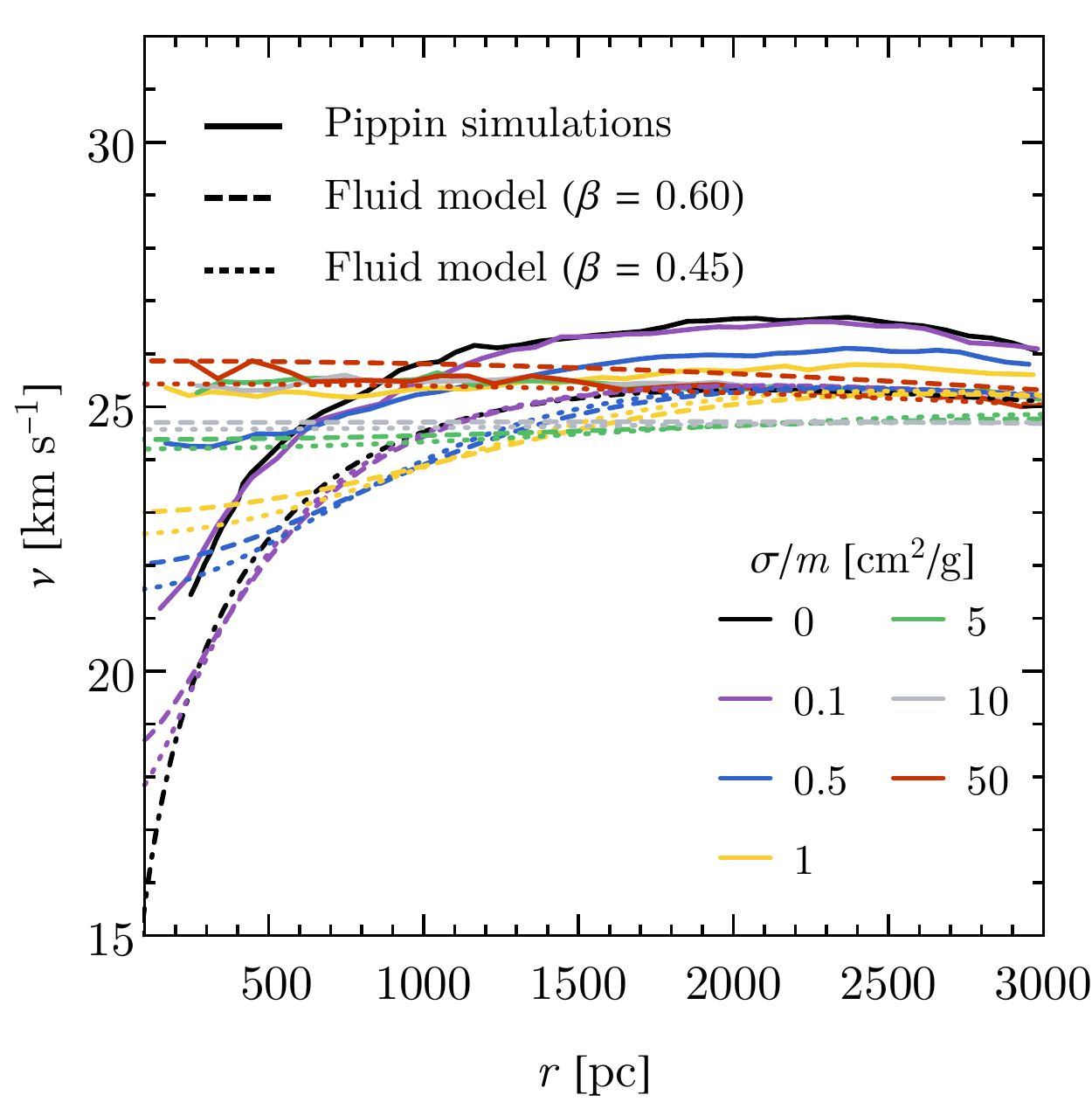}
   \caption{Comparisons of density (left) and 1D-velocity dispersion (right) profiles at the end of Pippin simulations~\cite{Elbert:2014bma} (solid) with results from fluid models (dotted and dashed) at $t = 10~{\rm Gyr}$.  Different colors indicates simulations with different values of $\sigma/m$. The dashed (dotted) lines are set with the calibration parameter $\beta = 0.60$ ($\beta = 0.45$). The black dot-dashed lines represent initial density and velocity dispersion profiles in the fluid simulations. The Pippin halo parameters are $r_s=2.7~{\rm kpc}$ and $\rho_s=1.73\times10^{7}~M_\odot/{\rm kpc^3}$. The difference between the cosmological $N$-body simulations and the fluid model is within a factor of 2 in the density profile comparison and within a factor of 1.2 in the 1D-velocity dispersion comparison.}
   \label{fig:pippin}
\end{figure*}

\begin{figure}[h]
   \centering
   \includegraphics[width=0.4\textwidth]{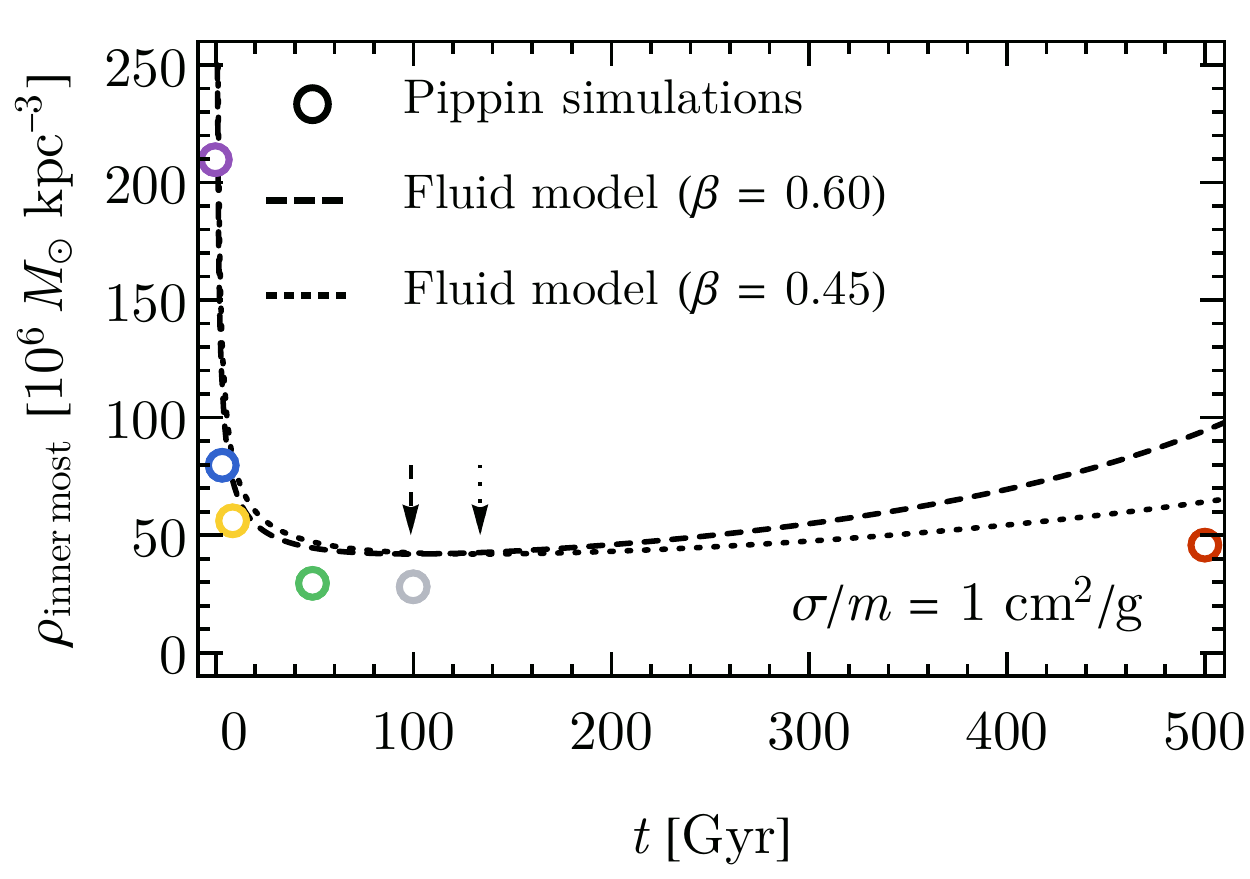}    \caption{Calibrating coefficient $\beta$ with the central density extracted from Pippin simulations. The central density for $\sigma/m= 1 \cmg$ at $1~\rm{Gyr}$ (purple), $5~\rm{Gyr}$ (blue), $10~\rm{Gyr}$ (yellow), $50~\rm{Gyr}$ (green), $100~\rm{Gyr}$ (gray), and $500~\rm{Gyr}$ (red) are extracted from Pippin density snapshot for $\sigma/m = 0.1, 0.5, 1, 5, 10$, and $50 \cmg$ at $10 ~{\rm Gyr}$. The dashed and dotted lines are represent fluid simulations with $\sigma/m = 1\cmg$ and calibrated with $\beta = 0.60$ and $\beta=0.45$ respectively. The downward arrow indicates the moment for stage $1 \to 2$ transition.}
   \label{fig:calibrationpippin}
\end{figure}

We take the calibrated $\beta$ ($0.59\leq \beta \leq 0.61$) and cross check the fluid simulation results with that of the cosmological $N$-body simulations in Ref.~\cite{Elbert:2014bma}. Ref.~\cite{Elbert:2014bma} showed the snapshots of density and 3D velocity dispersion profiles for today for a dwarf galaxy halo that they call ``Pippin''. They are shown as solid lines in the left and right panels of~\figref{pippin} (note that we translate the 3D-velocity dispersion to a 1D-velocity dispersion by $\nu = \nu^{\rm 3D}/\sqrt{3}$). The parameters for Pippin~\cite{Elbert:2014bma} are $M_{\rm vir}=9\times10^{9}M_\odot$, $r_s=2.7~{\rm kpc}$, and $\rho_s=1.73\times10^{7}~M_\odot/{\rm kpc^3}$, where the subscript ``vir" indicates that the halo boundaries are set by their virial radius. On the fluid simulation side, we assume the initial profile to be an NFW profile for $\sigma/m= 0$ today. We also assume the evolution time of the halo is $10~{\rm Gyr}$. We then take the Pippin halo setup and run the fluid simulations for $\sigma/m =0.1, 0.5, 1, 5, 10$, and $50\cmg$. The simulations are truncated at $10~{\rm Gyr}$ and snapshots for the density and velocity dispersion profiles with $\beta=0.60$ are shown as dashed lines in the left and right panel of~\figref{pippin}, respectively.  Varying $\beta$ to $0.59$ from $0.61$ does not bring noticeable changes to the snapshots. Generically, the fluid model overestimates the density and underestimates the velocity dispersion in comparison with the values provided by the cosmological $N$-body simulations. Nevertheless, the discrepancies in the density and velocity-dispersion snapshot are within a factor of $2$ and $1.2$ respectively. 

For $\sigma/m = 50~\cmg$, Pippin sees a ``moderate collapse". This is also observed in the fluid model, where we see the halo evolves close to the end of stage 2 at $t = 10~{\rm Gyr}$ with $\sigma/m = 50~\cmg$. Nevertheless, the Pippin halo exhibit a slower collapse than the fluid simulated halo with $\beta = 0.60$. To better see this, we notice that $\kappa_{\rm lmfp} \propto \sigma/m$, i.e., an SIDM halo with $\sigma/m = 5 \cmg$ evolves $5$ times faster than that of $\sigma/m = 1 \cmg$ during stage 2, the bulk of the evolution. This proportionality induces a degeneracy: the snapshot for a Pippin halo with $\sigma/m = 5 \cmg$ at $10~{\rm Gyr}$ is identical to that with $\sigma/m = 1 \cmg$ at $50~{\rm Gyr}$. Adopting this degeneracy implied by the fluid model, we translate the density snapshots for Pippin halo with $\sigma/m = 0.1, 0.5, 1, 5, 10$, and $50 \cmg$ at $t= 10~{\rm Gyr}$ into snapshots with $\sigma/m = 1\cmg$ at $t= 1, 5, 10, 50, 100$, and $500~{\rm Gyr}$, respectively. The averaged inner densities for each snapshot are shown as colored circles in~\figref{calibrationpippin}. In comparison with the fluid simulations with $\beta=0.60$ (dashed line), the Pippin simulations show a slower collapse between $100~{\rm Gyr}$ and $500~{\rm Gyr}$. 

\begin{figure*}[h]
   \centering
   \includegraphics[width=0.4\textwidth]{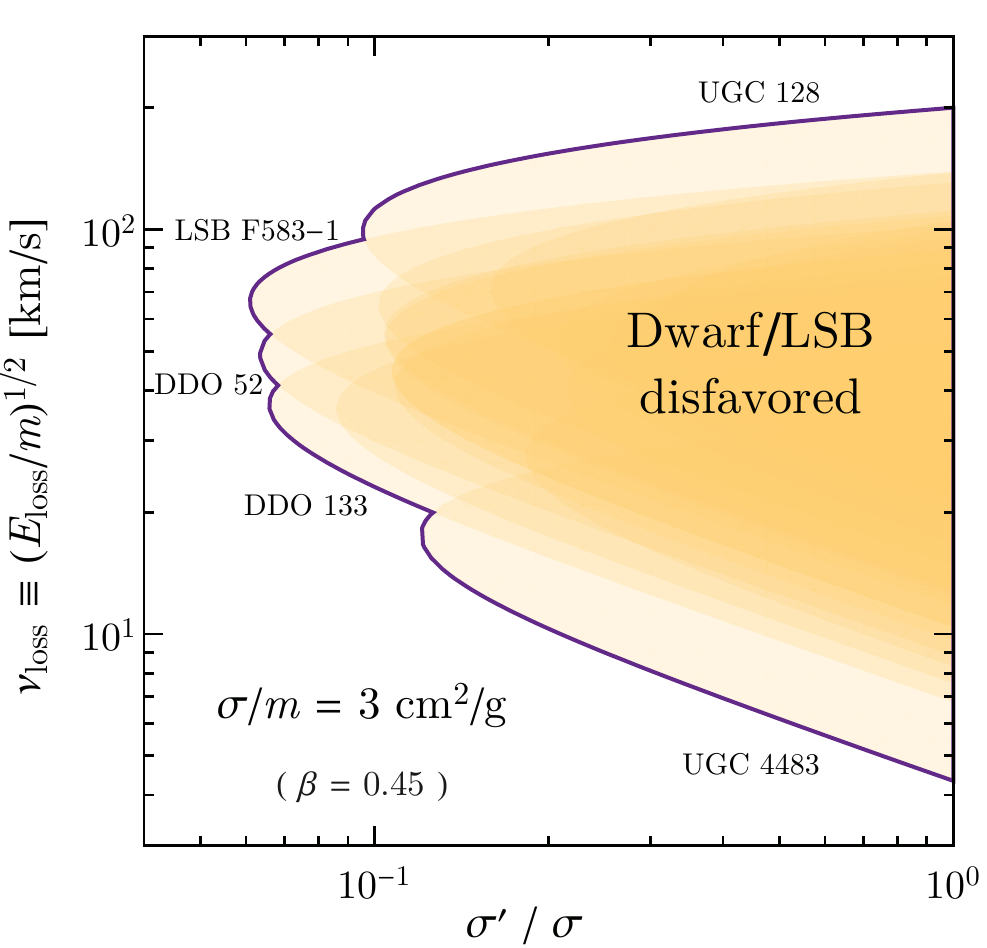}~\includegraphics[width=0.4\textwidth]{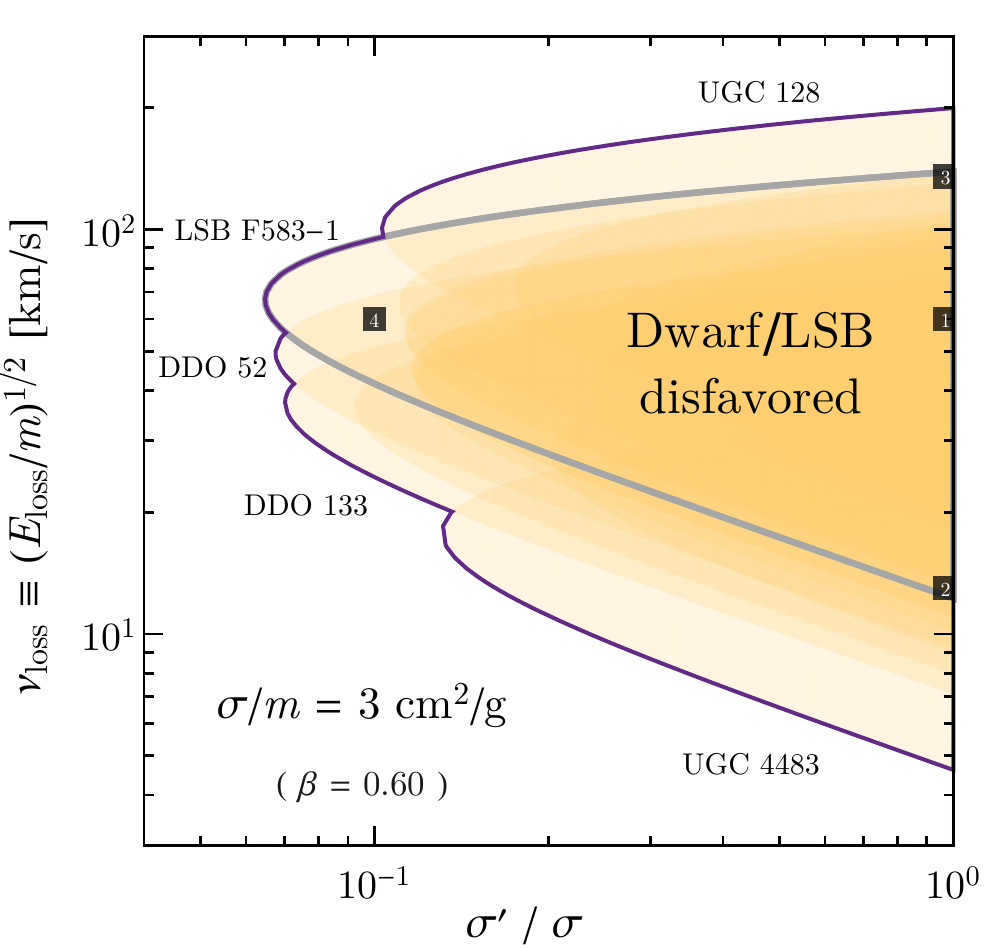}    \caption{Left: Constraints on the dissipative parameters from the absence of core collapse in individual dwarf galaxies (yellow) within $10~{\rm Gyr}$ and its overall boundary for the sample (purple). We take $\beta=0.45$. Right: Constraints for $\beta=0.60$. The panel is reproduced from Fig.~3 in the main text.}
   \label{fig:beta45}
\end{figure*}

This slower collapse is better captured by the fluid simulations slowed down by a factor about $4/3$ (dotted line), or equivalently, the fluid simulations with $\beta = 0.45$, since that a smaller $\beta$ implies a slower collapse. In~\figref{pippin}, we thus also show snapshots with $\beta = 0.45$ for various cross section strengths at $10~{\rm Gyr}$ as dotted lines. In the left panel of~\figref{beta45}, we show constraints on the dissipative parameters for $\beta=0.45$, which are similar to the results with $\beta=0.60$ as illustrated in Fig. 3 of the main text (reproduced as the right panel of~\figref{beta45}).

Our fluid simulations do not take into account mergers and environmental effects of the continuous infall of the background matter and the presence of baryons. A major merger can rebuild cuspy profiles and reset the evolution clock, delaying the thermal evolution. The self-similar accretion shock heating can prevent gravothermal collapse of an SIDM halo if the mass accretion rate is high~\cite{Ahn:2003xb}. However, cosmological $N$-body simulations show that both major merger and mass accretion rates decrease sharply towards low redshifts, and they become negligible for halos at present~\cite{Stewart:2008ep,McBride:2009ih}. This may explain the good agreement between the simulations and the semi-analytical estimates for the Pippin halo. Thus, we expect our results are robust for near-field galaxies at redshift $0$. Since our constraints on the dissipation parameters are based on the dwarf galaxies with low baryon content, baryons do not play an important role in setting the limits. If the halo's baryonic content is high, the baryons would speed up core collapse~\cite{Elbert:2016dbb,Sameie:2018chj}.

So far, we have focused on the calibration with simulations for purely-elastic dark matter self-interactions. Only very recently, Choquette et al. performed $N$-body simulations for dissipative SIDM~\cite{Choquette:2018lvq}. Their results confirm the overall predictions of our fluid model, i.e., dissipative dark matter self-interactions speed up the onset of core collapse, and a cuspy inner density profile emerges for a collapsed halo (as one can inferred from their mass profiles shown in their Fig.~3).

We take their Model B and make a detailed comparison. The models assumes the energy dissipation becomes active when the relative velocity of the incoming particles is greater than twice of the critical velocity $v_c$, while the cross section always keeps the same. We approximate the model by taking $\nu_\text{loss} = v_c$ and $\sigma'/\sigma=1$ for our simulations. In addition, we fix the halo parameters as in~\cite{Choquette:2018lvq}: an isolated halo with a total mass of $10^{11}\,M_\odot$ and an initial NFW profile with $\rho_s = 1.49\times 10^6\,M_\odot/\text{kpc}^3$ and $r_s = 11.1\,\text{kpc}$. The cross sections are fixed to be $\sigma'/m=\sigma/m = 38 \cmg$ ($\sigma'/m=\sigma/m= 0.13$) and the cooling parameters $\hat \nu_\text{loss}\equiv\nu_\text{loss}/\nu_0 = 0.12 $ and $\hat \nu_\text{loss} = 0.39$. From the fluid-model simulation with $\beta=0.60$, we find the time reduction for core collapse is $90$ for $\hat \nu_\text{loss}= 0.12$, which is a factor of $1.3$ larger than $69$, seen in their inelastic SIDM simulations. While, for $\hat \nu_\text{loss} = 0.39$, we get $600$, a factor of $3.6$ larger than their simulated value $166$~\cite{Choquette:2018lvq}. One possible reason for the discrepancy in the latter case is that the cooling effect is so strong that the hydrostatic equilibrium assumption may not hold exactly. In addition, some subtle differences between the two approaches may also play a role, as discussed in~\cite{Choquette:2018lvq}. Although our fluid model may overestimate the time reduction by an ${\cal O}(1)$ factor, compared to the $N$-body simulations, the bounds on $\nu_\text{loss}$ in Fig.~3 of the main text do not change much. For the regions close to the upper and lower limits of $\hat \nu_\text{loss}$, $\xi$ depends on $\hat \nu_\text{loss}$ exponentially, as indicated in the Fig.~2 (right) of the main text and explicitly shown in the equation near the end of Sec. III. Thus, an ${\cal O}(1)$ change in $\xi$ only leads to a minor adjustment in the limits of $\hat \nu_\text{loss}$. Thus, our fluid model calibrated with elastic simulations works well in capturing overall features of halo evolution in the presence of the dissipative interactions and our bounds our robust.

\stepcounter{sec}
\section{\Alph{sec}.~Data of Dwarf and LSB Galaxies}

In \tabref{final}, we list all dwarf/LSB galaxies with low baryon-content fitted in~\cite{Kamada:2016euw}. We translated the derivation (2nd column) on the concentration $c_{200}$ into the value of $c_{200}$ according to the concentration function used in~\cite{Kamada:2016euw} (taken from~\cite{Dutton:2014xda}).  The subscript ``$200$" indicates that the halo boundaries are set by the  radius where the averaged density is 200 times of the critical density of the Universe. Based on the concentration $c_{200}$ and total mass $M_{200}$, we determined $r_s$ and $\rho_s$ and further construct fiducial quantities listed in~\tabref{fiducial} for each halo.

\begin{table}[t]
   \centering
      \topcaption{Parameters of dark matter halos for $18$ dwarf/LSB galaxies with low baryon content used in Fig.~3 of the main text. The fits are taken from~\cite{Kamada:2016euw}.
      }
   \label{tab:final}
   \begin{tabular}{@{} lrrrrr @{}} 
      \hline
     Name  & $c_{200}$ & dev.~[$\sigma$] & $M_{200}\,[M_\odot]$  & $r_s$ [kpc] & $\rho_s\,[M_\odot\,{\rm kpc}^{-3}]$ \\
      \hline

UGC 4483	&	16.1& 0	&	$1.5\times 10^{9}$	&	1.5	&	$1.86\times 10^{7}$	\\
DDO 126	&	10.4& $-1$	&	$9\times 10^{9}$	&	4.2	&	$6.31\times 10^{6}$	\\
DDO 133	&	16.8& $1$	&	$1.2\times 10^{10}$	&	2.9	&	$2.08\times 10^{7}$	\\
DDO 154	&	14.7& $0.5$	&	$1.3\times 10^{10}$	&	3.4	&	$1.48\times 10^{7}$	\\
NGC 2366	&	12.2 & $0$	&	$2.3\times 10^{10}$	&	4.9	&	$9.33\times 10^{6}$	\\
UGCA 442	&	11.9 & $0$	&	$3\times 10^{10}$	&	5.5	&	$8.73\times 10^{6}$	\\
UGC 1281	&	11.9 & $0$	&	$3\times 10^{10}$	&	5.5	&	$8.73\times 10^{6}$	\\
DDO 52	&	15.3 & $1$	&	$3\times 10^{10}$	&	4.3	&	$1.65\times 10^{7}$	\\
DDO 87	&	8& $-1.5$	&	$3.5\times 10^{10}$	&	8.6	&	$3.33\times 10^{6}$	\\
NGC 3109	&	11.2 & $0$	&	$5.5\times 10^{10}$	&	7.2	&	$7.5\times 10^{6}$	\\
NGC 1560	&	11.1	& $0$ &	$6\times 10^{10}$	&	7.4	&	$7.34\times 10^{6}$	\\
LSB F583-1	&	13.9	& $1$ &	$8\times 10^{10}$	&	6.5	&	$1.28\times 10^{7}$	\\
UGC 5750	&	7.4	& $-1.5$ &	$8\times 10^{10}$	&	12.3	&	$2.72\times 10^{6}$	\\
UGC 3371	&	6.4	& $-2$	& $9\times 10^{10}$	&	14.7	&	$1.96\times 10^{6}$	\\
UGC 11707	&	10.5 & $0$	&	$1\times 10^{11}$	&	9.3	&	$6.47\times 10^{6}$	\\
IC 2574	&	5.4	& $-2.5$	& $1.5\times 10^{11}$	&	20.9	&	$1.3\times 10^{6}$	\\
UGC 5005	&	7.7	& $-1$	&$1.8\times 10^{11}$	&	15.5	&	$3.03\times 10^{6}$	\\
UGC 128	&	9.2	&	$0$ & $3.8\times 10^{11}$	&	16.6	&	$4.65\times 10^{6}$	\\

\hline

   \end{tabular}

\end{table}

\stepcounter{sec}
\section{\Alph{sec}.~Benchmarks for LSB F583-1}

\begin{table}[htbp]
   \centering
   \topcaption{
   Particle physics benchmarks for LSB F5831-1. Benchmark (a) represents pure elastic scattering without any dissipative interaction (i.e., without any cooling).}  \label{tab:LSBF583}
   \begin{tabular}{@{} crrrrc @{}} 
      \hline
  &    $\sigma/m$ [$\cmg$]    & $\sigma'/\sigma$ & $\nu_{\rm loss}$ [$\kms$] & $t_c (t'_c)$ [${\rm Gyr}$] & Collapse \\
      \hline
      a & 3 & -- & -- & $1.7\times 10^2$ & No\\

      b & 3   &  1 & 13 & 8.5 & Yes\\
      c & 3   &  1 & 135 & 7.6 & Yes\\
      d & 3   &  0.1 & 60 & 5.9 & Yes\\
      e & 3 &  1 & 60 & 0.22 & Yes\\

      \hline
   \end{tabular}
\end{table}

In Sec. IV,  we use LSB F5831-1 as an example and illustrate the estimates of the collapse time $t_c, t'_c$ for five particle physics benchmarks (summarized in~\tabref{LSBF583} for the reader's convenience). In~\figref{benchmarks}, we show the snapshots of the density evolutions for each benchmark to validate our $t_c$ estimation and illustrate some details of collapse with bulk cooling.  Here we assume $\beta=0.60$. Solid lines with various colors in~\figref{benchmarks} represent density snapshots at different times. The black dashed line shows the density profile inferred from the rotation curve fit. The 3D views of the density evolution of benchmark (a) and (b) are shown in the left panel of Fig.~1 in the main text.

\stepcounter{sec}
\section{\Alph{sec}.~Procedure for Setting Limits on Dark Matter Parameter Space}
\label{sec:procedure}

In this subsection, we discuss our procedure for determining the excluded region of Fig.~3 in the main text in further detail.

Our criterion for excluding a point in the dissipative dark matter parameter space is equivalent to ruling out those parameters for which the dwarf and LSB halos reach stage 3 of their gravothermal evolution in less than $10~{\rm Gyr}$. This criterion is conservative, since it requires that a cusp has been reestablished through the entire inner core region. Before our criterion is achieved, several other deviations from the cored profile may become apparent. For example, the central density of the halo will begin to exceed its minimum value well before the runaway collapse of stage 3 begins. Likewise, the log-slope of the density becomes steeper than the cored prediction at some point during stage 2.

\begin{figure*}[!htbp]
{\includegraphics[width=0.33\textwidth]{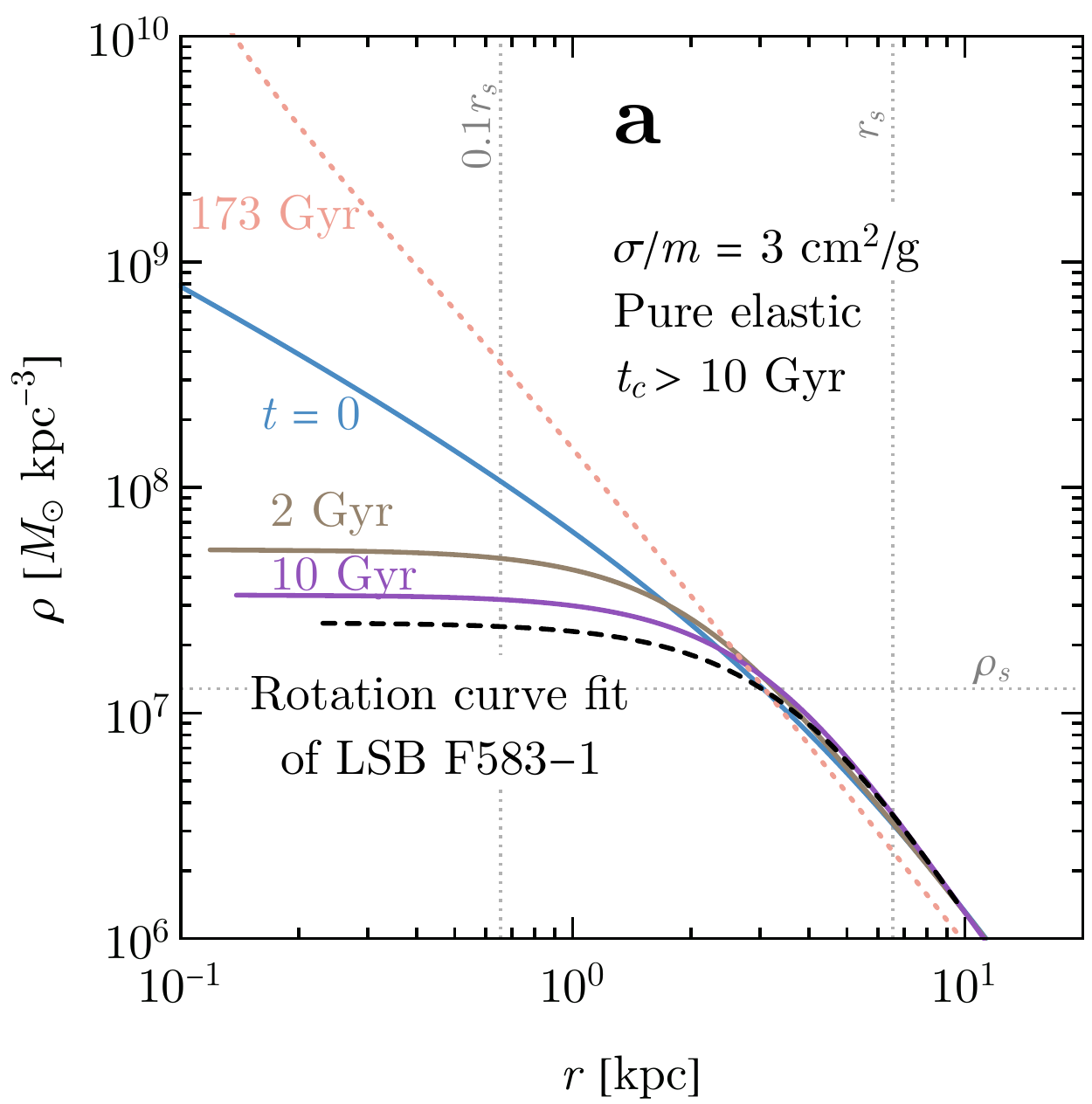}}~
{\includegraphics[width=0.33\textwidth]{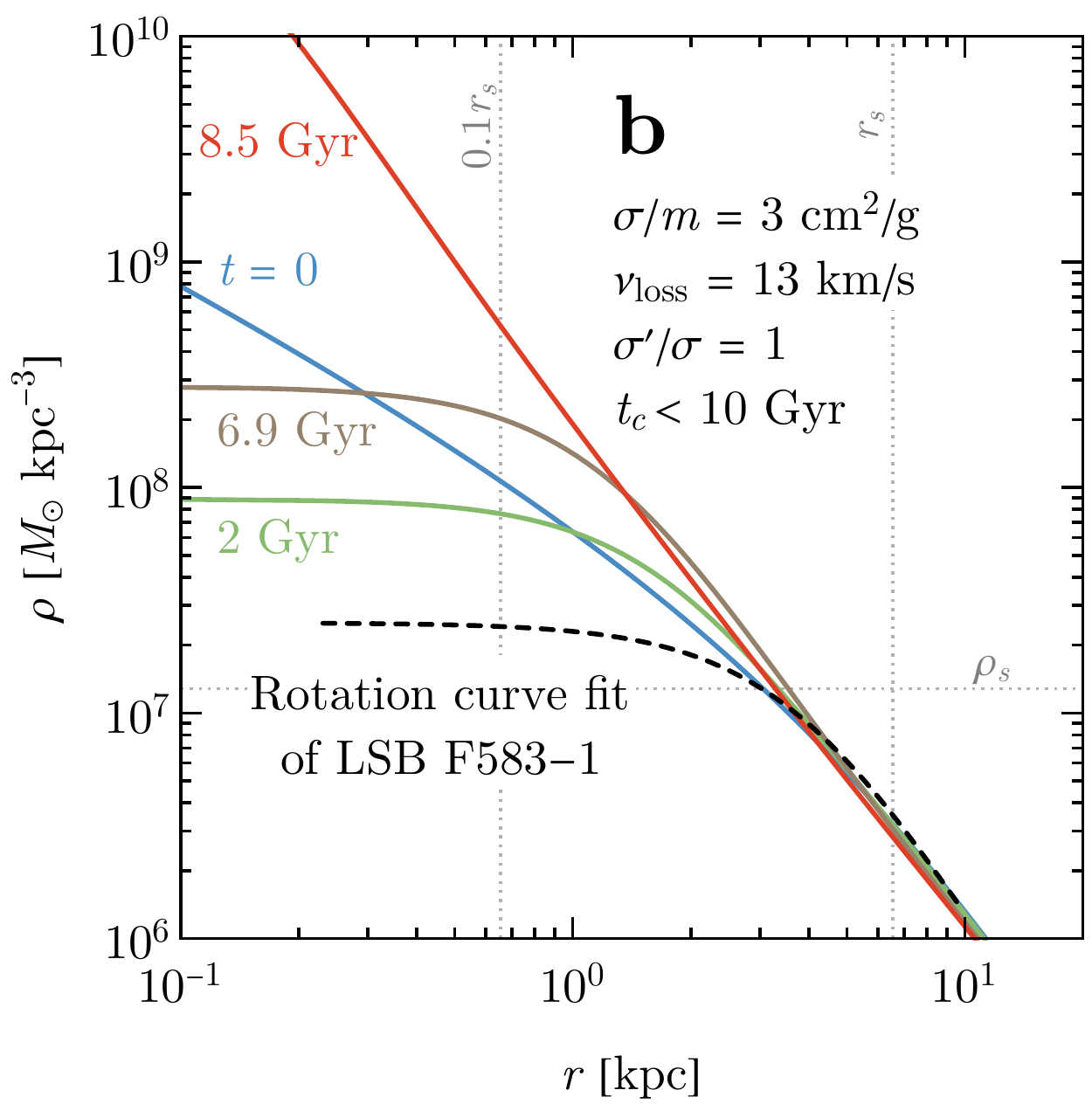}}~
{\includegraphics[width=0.33\textwidth]{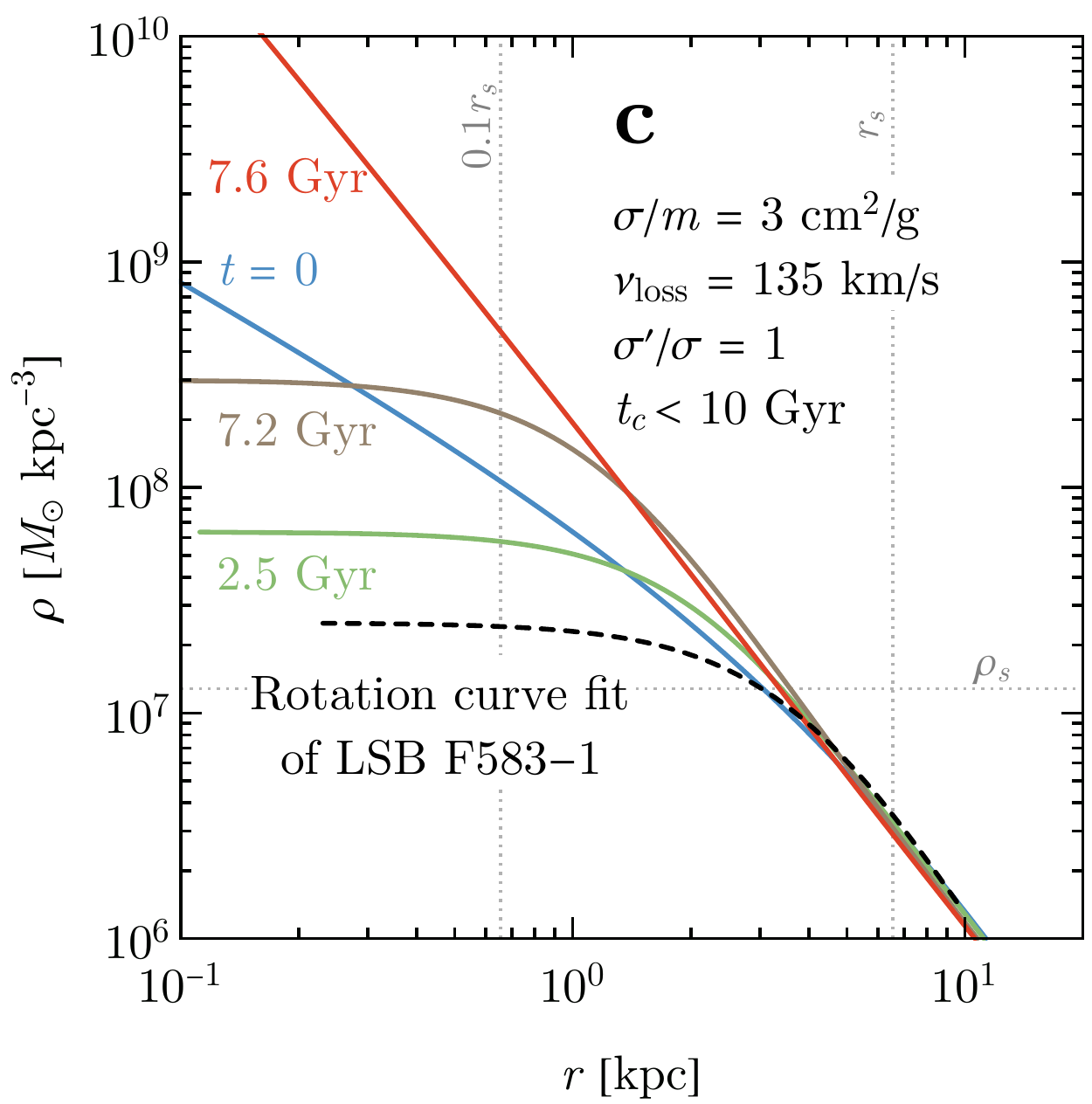}}\\
{\includegraphics[width=0.33\textwidth]{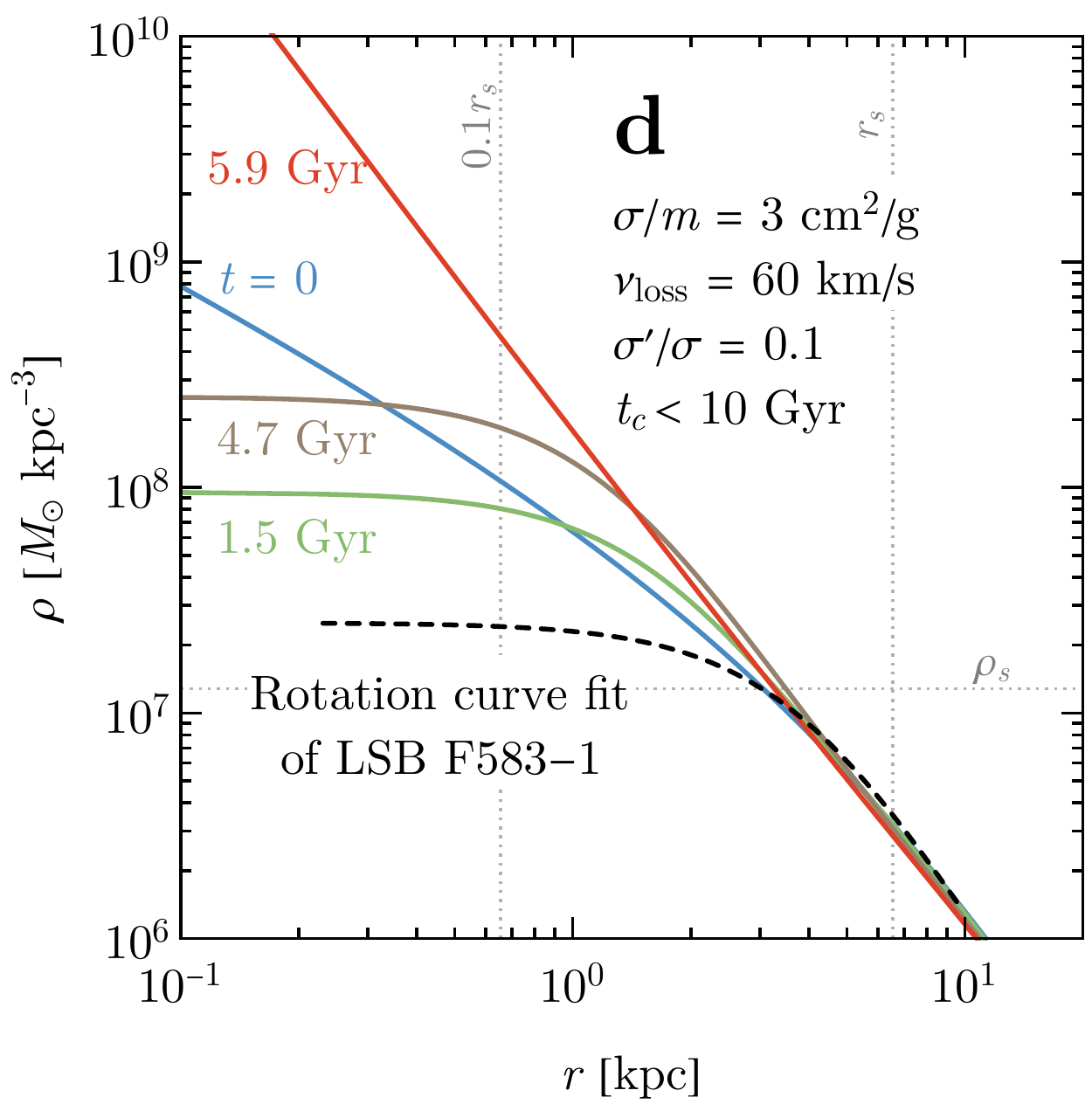}}~
{\includegraphics[width=0.33\textwidth]{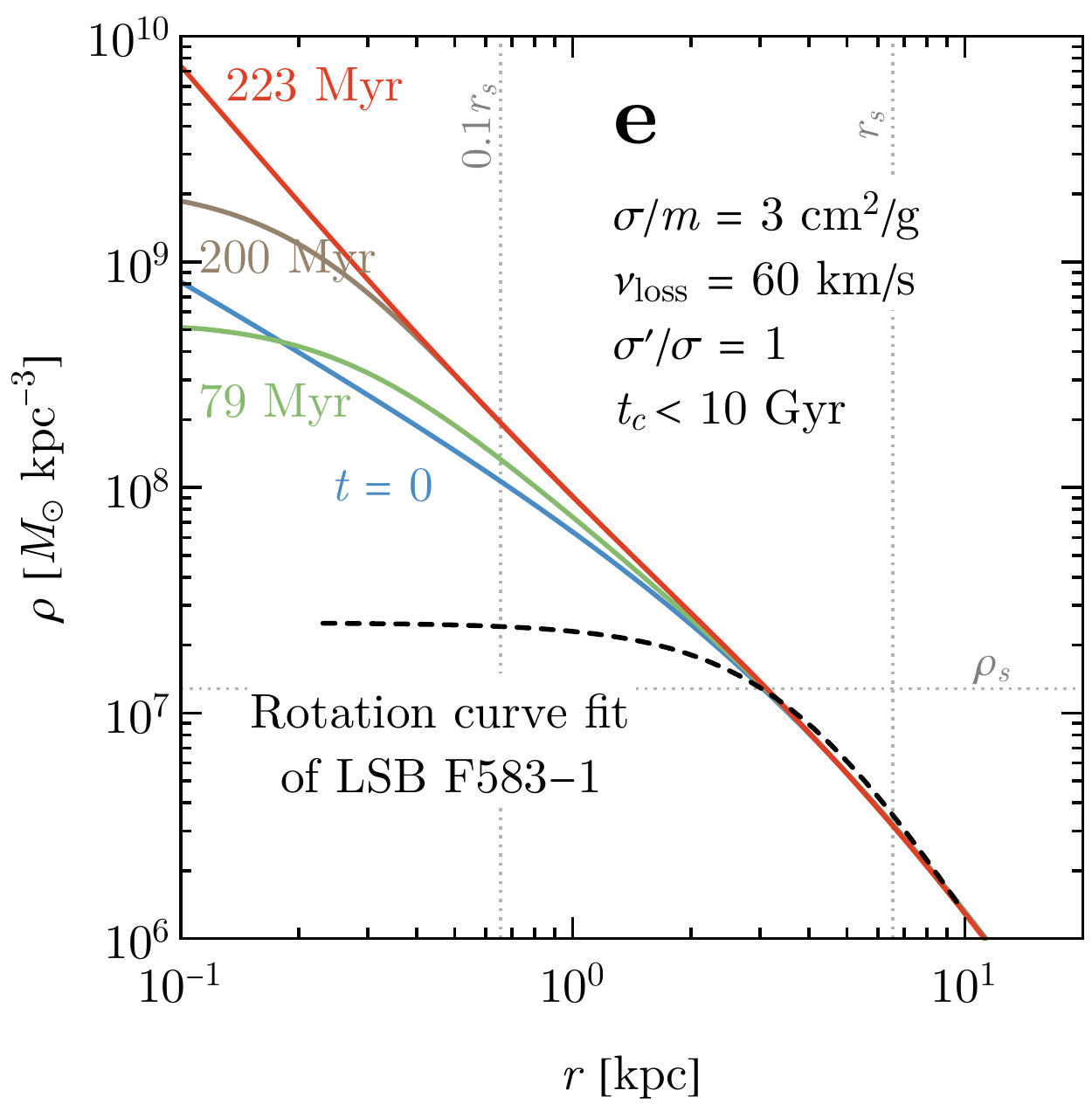}}
\caption{Evolution of the density profile for five dissipative dark matter benchmark points (a--e) listed in Sec IV (summarized in~\tabref{LSBF583}) for LSB F583-1 halo. We choose $\beta = 0.60$. Solid lines with various colors represent density snapshots at  various times. The blue solid lines indicate the initial NFW profile. The green solid lines correspond to the transition between stage 1 and 2, $t= t_{1\to 2}$. The red solid lines represent $t= t_c$. In the first panel, the red dotted lines illustrate the density profiles at $t = t_c > 10~{\rm Gyr}$. The black dashed lines indicate the density profile inferred from the rotation curve fit. }
   \label{fig:benchmarks}
\end{figure*}

However, these alternative criteria are somewhat more difficult to observe rigorously. As discussed in Sec.~D, the core density from a fluid simulation is up to a factor of 2 greater than that from the cosmological $N$-body simulation. We should regard this difference as a systematic  uncertainty  of the fluid model. To claim a robust exclusion based on the central density, a very large difference in the central density compared to the cored profile is necessary.

To illustrate our exclusion criterion, consider again~\figref{benchmarks}. As seen from panel (a), when the cooling is absent (benchmark (a)), the evolution already develops a large core around $2~{\rm Gyr}$ for $\sigma/m = 3\cmg$. Collapse eventually happens, but takes about $173~{\rm Gyr}$.  
The density profile is almost static for most of the halo's existence. The prediction of the central density of the fluid model is about $1.3$ times larger than the inferred density profile from the rotation curve fit of LSB F583-1, but this difference should be tolerated given the uncertainties in the rotation curve measurement and the systematic uncertainties in the fluid model.

Next we look at panels (b) and (c). The corresponding benchmark particle physics parameters (b) and (c) of \tabref{LSBF583} are inside the excluded region for LSB F583-1 but near the boundary, as shown in Fig.~3 of the main text. As is apparent in these panels,  the inner core density is already $\gtrsim 10$ times larger than from the fit at an age of $\sim 7~{\rm Gyr}$. After the halo is $\sim 8~{\rm Gyr}$ old, the entire inner halo is cuspy and the evolution reaches $t_c$. It is also worth noting that the core density (core size) at the maximal expansion is higher (smaller) than that of benchmark (a). This is because at stage 1, bulk cooling counters the inward heat flow, and hence limits the development of the core.  As a result, the particles in the inner halo experience fewer collisions during stage 1. This yields a smaller core with a higher core density. 

As we go deep into the excluded parameter space, the collapse time $t_c$ becomes even smaller, and the core size (density) at the maximal expansion becomes even smaller (higher) as shown in panel (d). Finally, when we reach benchmark (e) (panel (e)), we are near the strong cooling region and the collapse time is $223~{\rm Myr}$, which is near the free-fall time for the inner core. As shown in panel (e), the self-interactions never produce a large core beyond $0.1 r_s$. This yields an obvious contradiction with the observed density profile.

\end{document}